\documentclass[12pt,preprint]{aastex}

\newcommand{\ch}{{\it Chandra }}
\newcommand{\ulg}{ULIRGs}
\newcommand{\ul}{ULIRG}
\newcommand{\irof}{IRAS~05189-2524}
\newcommand{\iros}{IRAS~17208-0014}
\newcommand{\irto}{IRAS~20551-4250}
\newcommand{\irtt}{IRAS~23128-5919}
\newcommand{\ugc}{UGC~05101}

\newcommand{\chsq}{$\chi^2$}
\newcommand{\feka}{$Fe-K\alpha$}
\newcommand{\afe}{$\alpha/\rm{Fe}$}
\newcommand{\alphe}{$\alpha$-element}
\newcommand{\ha}{$\rm{H}\-\alpha$}

\slugcomment{Accepted for publication in ApJ}

\shorttitle{The Violent ISM: Dwarf Starbursts to ULIRGs}
\shortauthors{J Grimes et al.}

\begin{document}


\title{A {\it Chandra } X-ray Investigation of the Violent Interstellar
Medium: From Dwarf Starbursts to Ultraluminous Infrared Galaxies}


\author{J. P. Grimes, T. Heckman,
          D. Strickland, and A. Ptak}
\affil{Center for Astrophysical Sciences, Johns Hopkins University,
    Baltimore, MD 21218-2695}




\begin{abstract}
We have analyzed observations with the Chandra X-ray Observatory of the
diffuse emission by hot gas in 7 dwarf starburst galaxies,
6 edge-on starburst galaxies, and 9 Ultra Luminous Infrared Galaxies.
These systems cover ranges of $\sim10^4$ in
X-ray luminosity and several thousand
in star formation rate and K-band luminosity (a proxy for stellar mass).
Despite this range in fundamental
parameters, we find that the properties
of the diffuse X-ray emission are very similar in all three classes of
starburst galaxies. The spectrum of the diffuse emission is well fit
by thermal emission from gas with $kT \sim$ 0.25 to 0.8 keV and with
several-times-solar abundance ratios of $\alpha$ elements to Fe. The
ratio of the thermal X-ray to far-infrared luminosity is roughly  
constant,
as is the characteristic surface brightness of the diffuse X-ray  
emission.
The size of the diffuse X-ray source increases systematically
with both far-infrared and K-Band luminosity. All three classes
show strong morphological relationships between the regions of
hot gas probed by the diffuse X-ray emission and the warm
gas probed by optical line emission.
These findings suggest that the same physical mechanism is
producing the diffuse X-ray emission in the three types
of starbursts, and are consistent with that mechanism being
shocks driven by a galactic ``superwind'' powered by the kinetic
energy collectively supplied by stellar winds and supernovae in
the starburst.

\end{abstract}




\section{Introduction}

Observations of starburst galaxies have found that relatively bright
diffuse soft X-ray emission is always present
\citep[e.g.][and references therein]{strick04a}.
These regions of hot ($\rm{T\gtrsim10^6K}$) gas have been observed to
extend far beyond the starburst itself. In starbursts located in edge-on
disk galaxies, the emission can be traced many kpc out into the
galactic halo.  This diffuse X- ray emission is thought to be a prime  
manifestation of
the effects of ``feedback'' from star formation.
Short lived massive stars in the starburst inject kinetic energy
and highly metal-enriched gas into their surroundings through stellar
winds and supernovae. The collective effect is to drive a galaxy-scale
outflow known as a ``superwind'' \citep{heck90} out
of the starburst and into the galaxy halo and perhaps beyond.

These outflows could play a major role in the evolution
of galaxies and the intergalactic medium (IGM). For example,
by propelling metals out of
star-forming galaxies, they could explain both the
mass-metallicity relation of galaxies \citep[e.g.][]{trem04}
and the presence of substantial amounts of metals in the
intergalactic medium  \citep[e.g.][]{adel03,sav02} and
in the intracluster medium \citep[e.g.][]{loew04}.

Spectroscopy in the rest-frame ultraviolet has shown that
starburst-driven outflows are a characteristic feature of
starburst galaxies at high redshift
\citep{shap03,smail03}. However, the best laboratories
for studying the astrophysics of these winds are in local galaxies in
which their complex multi-phase nature can
be investigated in detail in the X-ray, ultraviolet, optical,
and radio regimes \citep[e.g.][2004]{hoop03}.
X-ray observations with high spatial resolution
are particularly crucial, as they most directly trace the hot
gas that powers the wind.

In recognition of this, the {\it Chandra} X-ray Observatory has devoted
significant amounts of time to the observation of starburst
galaxies and their winds. The galaxies studied span a broad range
from dwarf starbursts \citep[e.g.][]{mart02,hart04},
to edge-on disk galaxies \citep[e.g.][b]{strick04a}, to
interacting and merging systems \citep{fabb01,zez03}, to the Ultra  
Luminous Infrared Galaxies \citep[\ulg,][]{mcdo03,gall02,xia02}.
Halos of diffuse X-ray emission are seen in all these types
of starburst, and  in many cases the properties of the hot gas have  
been shown to be consistent with the superwind model. However,
investigations to date have mostly been of individual starburst
systems, or (in few cases) of small samples of similar
objects.

The time is now ripe for an investigation that exploits
the {\it Chandra} archive to uniformly
analyze the properties of the diffuse X-ray emission
in starburst galaxies spanning the widest possible range in the
properties of both the starburst itself and of its ``host'' galaxy.
In the present paper, we use \ch ACIS-S data to study
the impact of starbursts in three distinct samples of starbursts:
seven dwarf starbursts, six edge-on starburst galaxies
of intermediate luminosity, and nine ULIRGs.
By combining the data from the three galaxy types we cover a range of  
roughly $10^4$ in X-ray luminosity, and several thousand
in star formation rate and K-Band luminosity (a proxy for stellar mass).

Note that the edge-on starburst sample has been previously examined
extensively in two papers by \citet[b]{strick04a}. The properties
of the nuclear X-ray emission in the ULIRGs have been discussed
in \citet{ptak03}.

\section{The Data}

\subsection{Sample Selection}

Each of the three samples was selected to be complete
and unbiased. The ULIRG sample consists of the
eight nearest such objects ($cz <$ 15,000 km/sec). Following
\citet{ptak03} we have additionally included NGC 6240,
whose far-IR luminosity is just below the ULIRG range.
The members of the edge-on sample were selected to
be the nearest (closer than 20 Mpc) starbursts of intermediate
far-IR luminosity ($\sim$ few $\times 10^{10} L_{\odot}$)
in moderately massive disk galaxies
($\rm{M_{\ast}} \sim 10^{10}$ to $10^{11}  \rm{M}_{\odot}$)
viewed nearly edge-on (inclinations greater than 60 degrees).
This range in luminosity and mass is representative of typical
infrared selected starbursts in the present universe. Finally, 
our sample of
dwarf starbursts includes the five brightest (sum of ultraviolet
plus far infrared flux) starbursts in low mass galaxies
($\rm{M_{\ast}}$ below $\sim10^9 \rm{M}_{\odot}$). We have added two  
similar
galaxies He 2-10 and NGC 3077. The samples are listed
along with their most salient quantities in Table \ref{lumin}.

\subsection{Overview}

The X-ray data for our sample is based on \ch
ACIS-S observations of 22 galaxies. In order to obtain
a consistent data sample we have reanalyzed these observations
for all of the \ulg~and all but one of the dwarf starbursts.
For the edge-on starbursts we have used the luminosities, temperatures,
and images derived in \citet{strick04a}.  These were created in manner
similar to that described below for our analysis.  The data and analysis
for the dwarf starburst NGC 4214 was provided by \citet{hart04}.
From their background subtracted image and exposure maps
we have calculated the appropriate surface brightness and radial
values.  We have also used their derived flux and mekal temperature.
Exposure times, pointing information, and observation dates for all of
the galaxies are summarized in Table \ref{obsdata}.

For our analysis, data reprocessing and
flare removal was done as described in the CIAO threads from the CXC.
CIAO 3.01 with CALDB 2.23 was used
throughout the processing and analysis.
The two observations of \irof~were combined using the CIAO
task {\rm REPROJECT\_EVENTS}.  As the two observations are only
separated by 3 months we have ignored differences in
time dependent calibrations
such as the ACIS optical blocking filter contamination.  Only data
from the S3 chip has been included so we have not
attempted any charge transfer inefficiency corrections.

To isolate the central sources in the \ulg~
we have split each observation
into an inner (nuclear) and outer (halo) region.
In \citet{ptak03} 2-D elliptical gaussians
were fit to the inner source of each \ul.  We have doubled
the minor and major axis of the elliptical gaussian fits
and used that to define our inner region (Table \ref{innerregion}).
The outer regions
were manually chosen to include the rest of the galaxies extent (from
broad band smoothed images) and exclude the central region.
Extraneous point sources
detected by {\rm WAVDETECT} are excluded from the regions
during spectral extraction.
For the dwarf starbursts we have defined a single region that
includes all the X-ray emission from the galaxy and excludes all
point sources.

\subsection{Background subtraction}

Accurate background subtraction is important
when studying diffuse X-ray emission.
Spatial, spectral, and time variations in the
X-ray background complicate background analysis.
Therefore a region enclosing the entire S3 chip but excluding
extended and point sources was defined for each observation.
The count rate in these
regions was then calculated over several energy
bands  and for the appropriate CXC-provided
blank field datasets \citep{mark01}.
If the count rates and spectral shapes were similar
($<15\%$ change) the blank field background
was renormalized by the ratio of the background rates
in the 4-7 keV range.  The renormalized blank sky
background was then used for both the
spectral and image analysis.

The background rates for \irof, \iros, \irtt, MKN231, NGC 1705, and
NGC 5253 were all above normal quiescent levels even after
solar flare removal.  Several
of these observations were affected by extreme
solar activity which multiply impacted Chandra observations
during October 2001.  Due to the background spectral changes
during periods of high solar activity it is inappropriate to rescale the
quiescent blank field background datasets.  We have therefore used
local backgrounds for these observations.  The local backgrounds
are centered on the galaxies but exclude sources and the diffuse
emission around the galaxy. This local background was then used
for spectral background subtraction.  The image background however
was created by rescaling the blank field background images as described
above for the galaxies with quiescent background levels.

\subsection{Image Analysis}

\begin{figure}
\includegraphics[width=6in,keepaspectratio=true,clip=true,origin=bl]{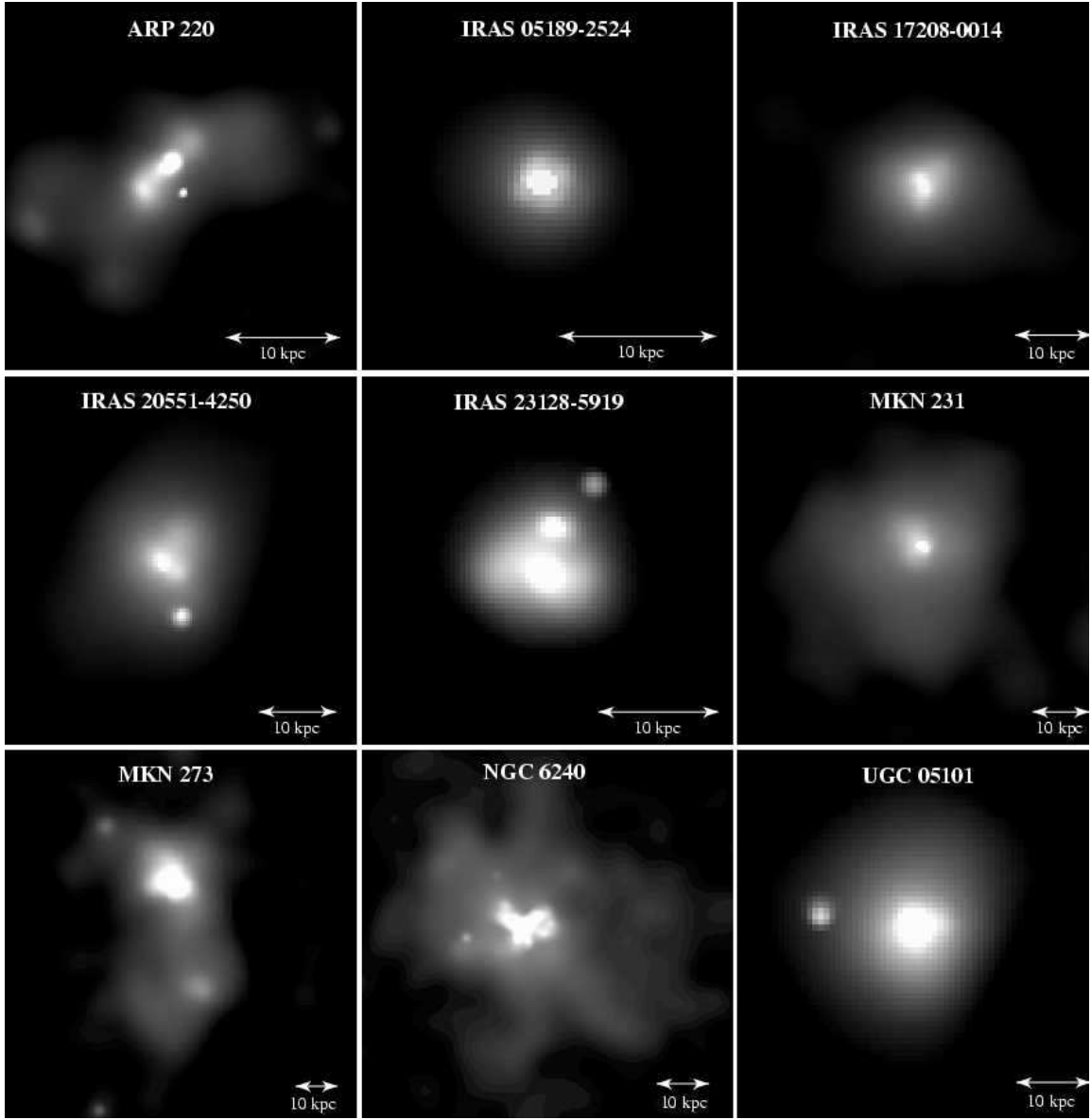}
\caption{\ulg:  Adaptively Smoothed Images of all 9 \ulg~in the 0.3-8.0
keV energy band.
A line of length 10 kpc has been drawn on each image.\label{smoothulg}}
\end{figure}

\begin{figure}
\includegraphics[width=6in,keepaspectratio=true,clip=true,origin=bl]{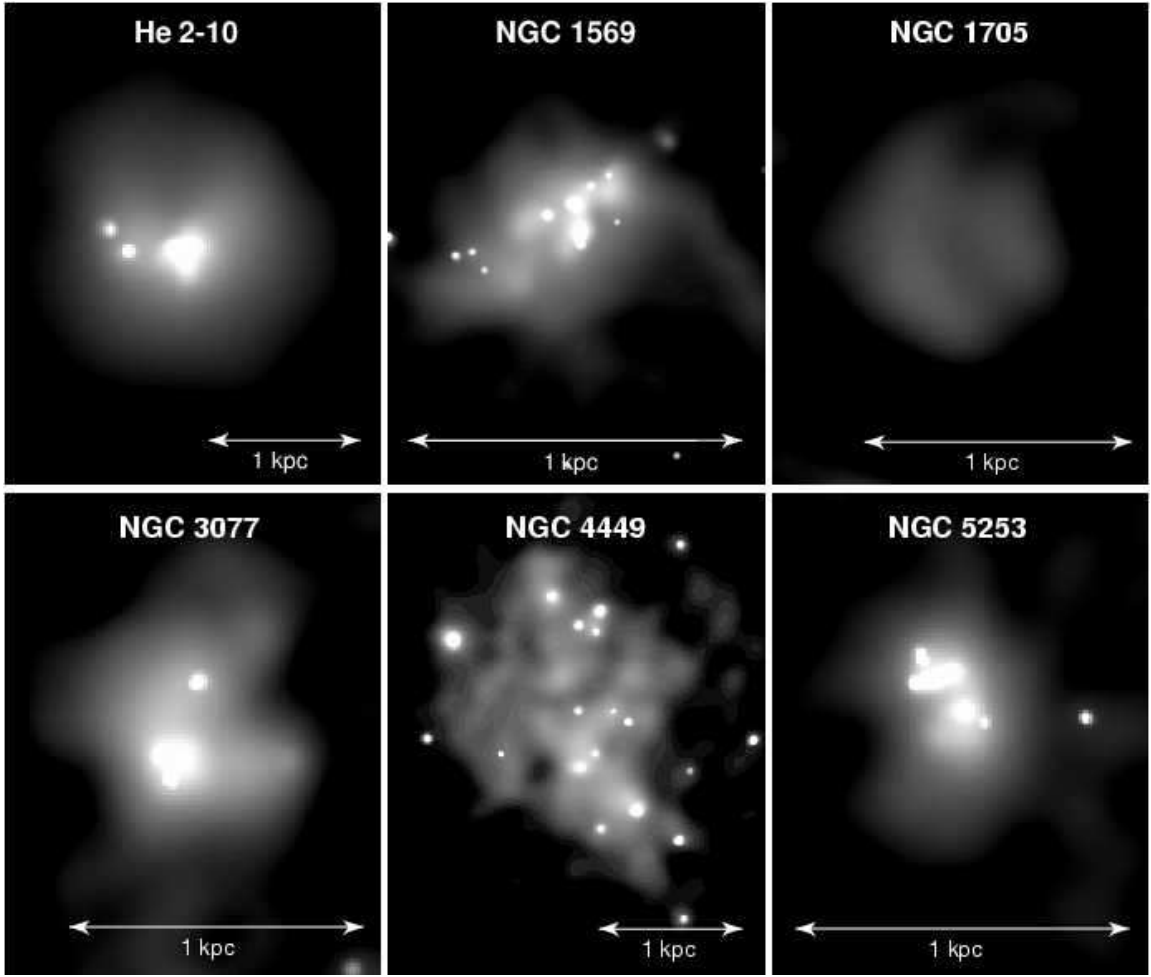}
\caption{Dwarf Starbursts:  Adaptively Smoothed Images of all 6 Dwarf
Starbursts in the 0.3-8.0 keV energy band.
A line of length 1 kpc has been drawn on each image.  These and the \ul~
images (Figure \ref{smoothulg}) show morphological similarities although
they vary greatly in their physical dimensions.
\label{smoothdwarf}}
\end{figure}

Using the renormalized quiescent background, background subtracted 
images were produced for every object in the sample.  The images
were created in a variety of energy bands
and then adaptively smoothed
using the {\rm CIAO} task {\rm CSMOOTH}.
Adaptively smoothed images for all of the
\ulg~and dwarf starbursts in the 0.3-8 keV
energy band can be found in Figures \ref{smoothulg} and
\ref{smoothdwarf}.  In order to create
representative color images for every galaxy we first
divided the 0.3 - 1.0 keV,
1.0 - 2.0 keV, and 2.0 - 8.0 keV images
by the monoenergetic exposure map for each observation.  The images
were adaptively smoothed
and combined into a single representative color image.

To derive the radial extent of the galaxies we
focused on the 0.3 - 1.0 keV background subtracted
image. This energy range was chosen as it
is dominated by the diffuse thermal emission.
Circular annuli starting at the
galaxy center and radiating outward to the end of the
diffuse X-ray emission were created for each observation.
Excluding all point sources except the central source, the number of
counts in each annulus
was extracted.  The 50\%, 75\%, 90\%, and 95\% counts enclosed
radii were then calculated (Table \ref{obsradii}).
The same annular regions were also used on an exposure map
corrected image of the galaxy to calculate the surface brightness
at the various radii.  An isometric radius was defined
as the radius that the surface brightness was
$1\times10^{-9}~\rm{counts/s/cm^2}$.  The AGN
\ul~\irof~is essentially pointlike in the Chandra
observations.  Therefore we are unable to determine 
\irof 's spatial extent and have not used it in our analyses
of galaxy size and surface brightness.

In the previous analysis of edge-on starburst
galaxies \citep{strick04a}, distances and surface brightness
computations are computed
separately for both directions along the major and minor axis.
As the other galaxies in our sample represent a variety of orientations
we have reanalyzed the edge-on starburst data using the same simple
radial geometry we used in analyzing the \ulg~and dwarf starbursts  
above.

\subsection{Spectral Fitting}

Spectra were extracted for all of our defined regions.
After background subtraction
the spectra were rebinned to a minimum of 20
counts per bin to allow use of the \chsq~statistic.
For every fit we have included
Galactic absorption and an ACIS contamination
model \citep{char02}.  The Galactic absorber column was
fixed to the value obtained from COLDEN using the NRAO H~I
dataset and found in Table \ref{obsdata} \citep{dick90}.  
Allowing the absorption column to
vary did not significantly improve fits and generally
was consistent within one sigma of the COLDEN value.

\begin{figure}
\centering
\leavevmode
\columnwidth=.40\columnwidth
\includegraphics[width=1.8in,angle=-90]{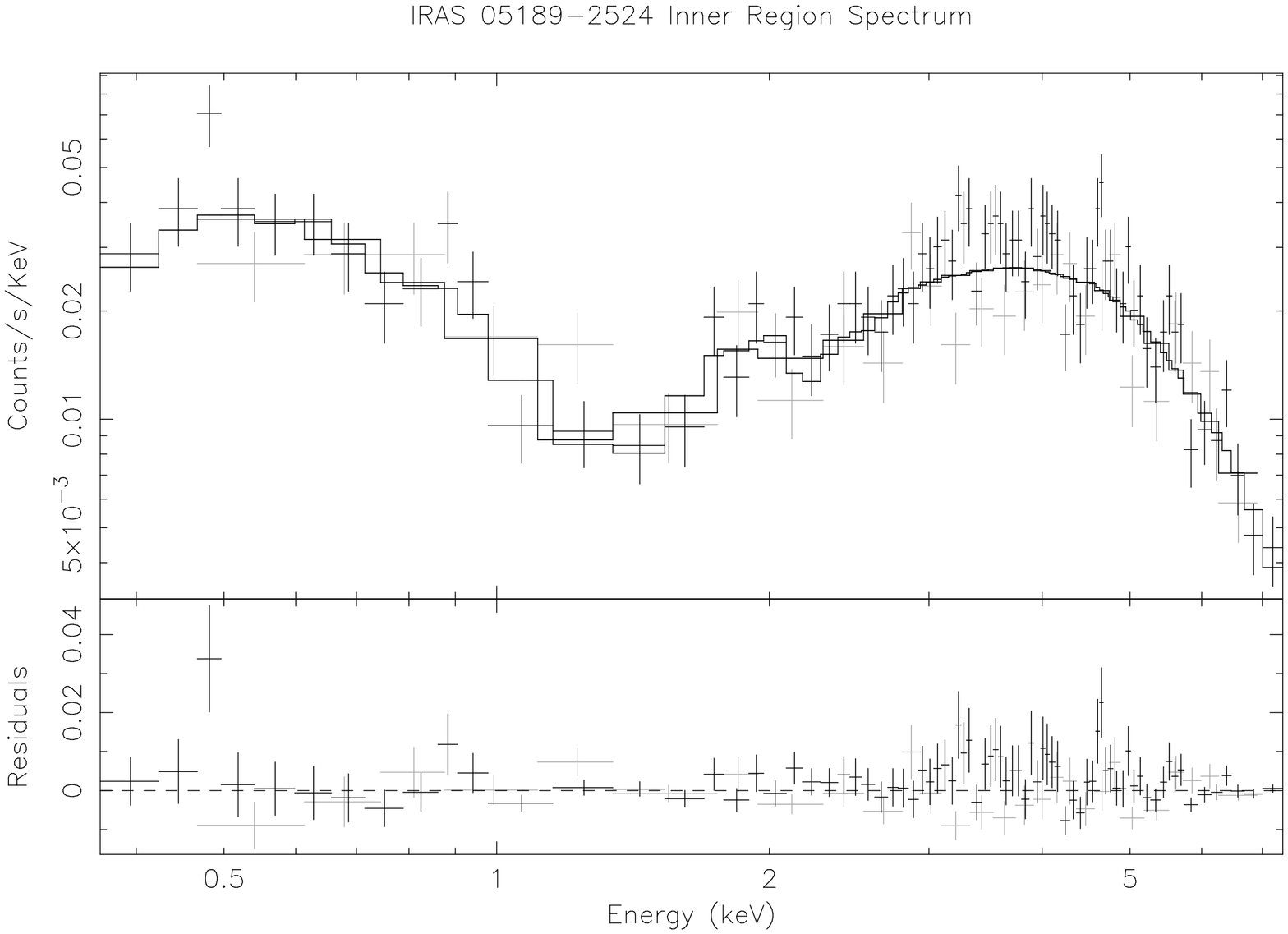}
\hfil
\includegraphics[width=1.8in,angle=-90]{f3b.eps}
\hfil
\vspace{.00in}
\includegraphics[width=1.8in,angle=-90]{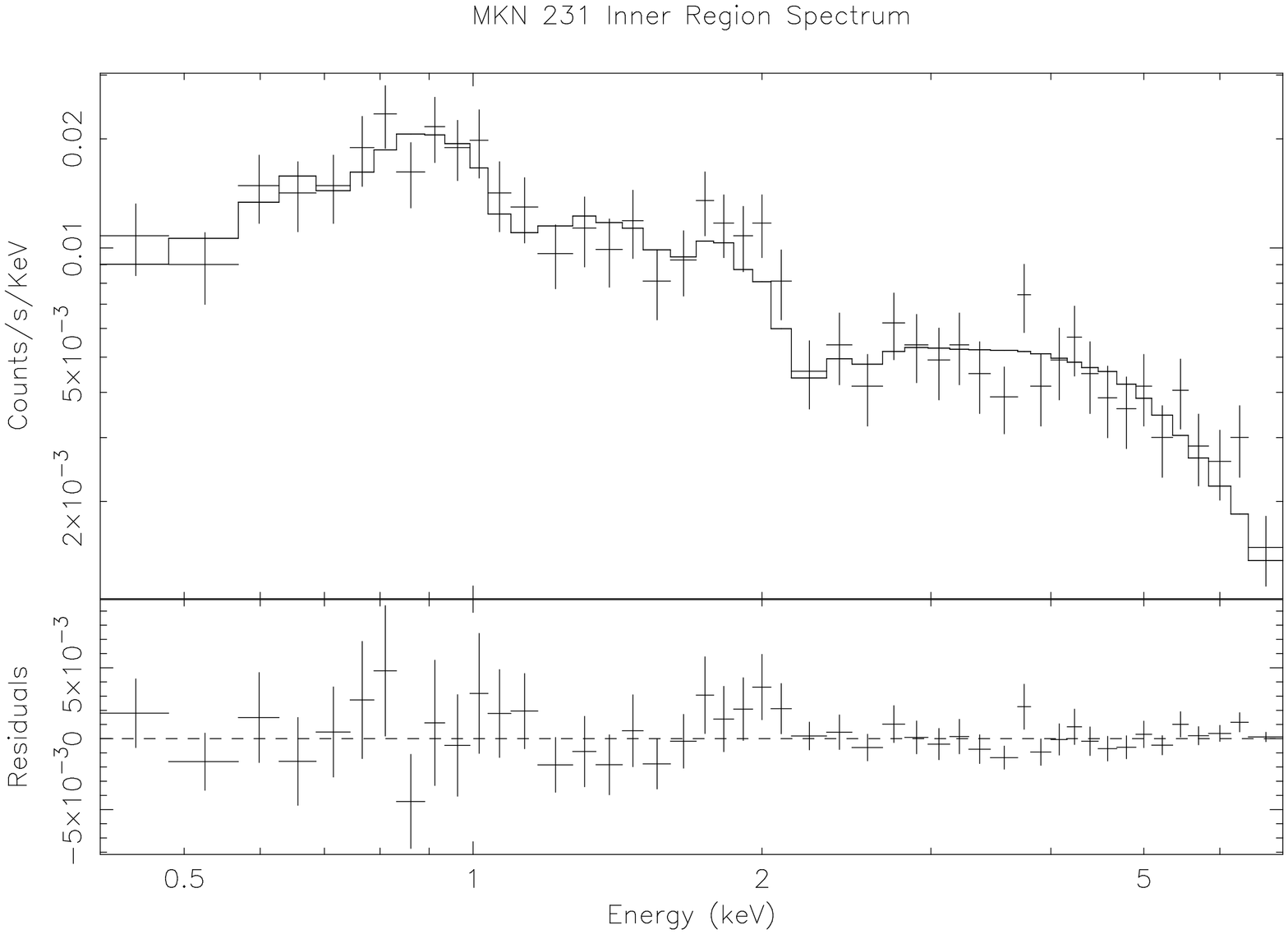}
\hfil
\includegraphics[width=1.8in,angle=-90]{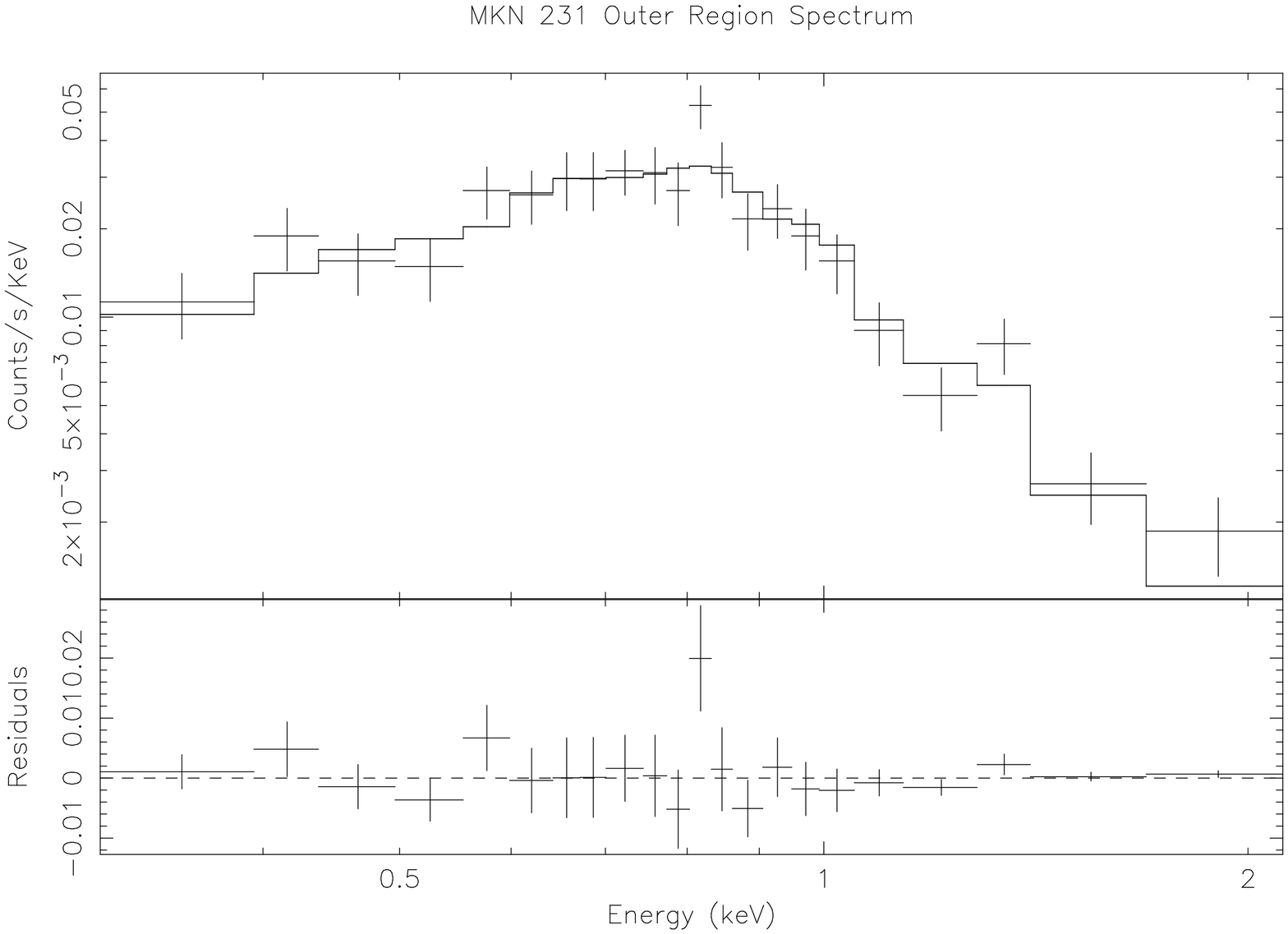}
\hfil
\vspace{.00in}
\includegraphics[width=1.8in,angle=-90]{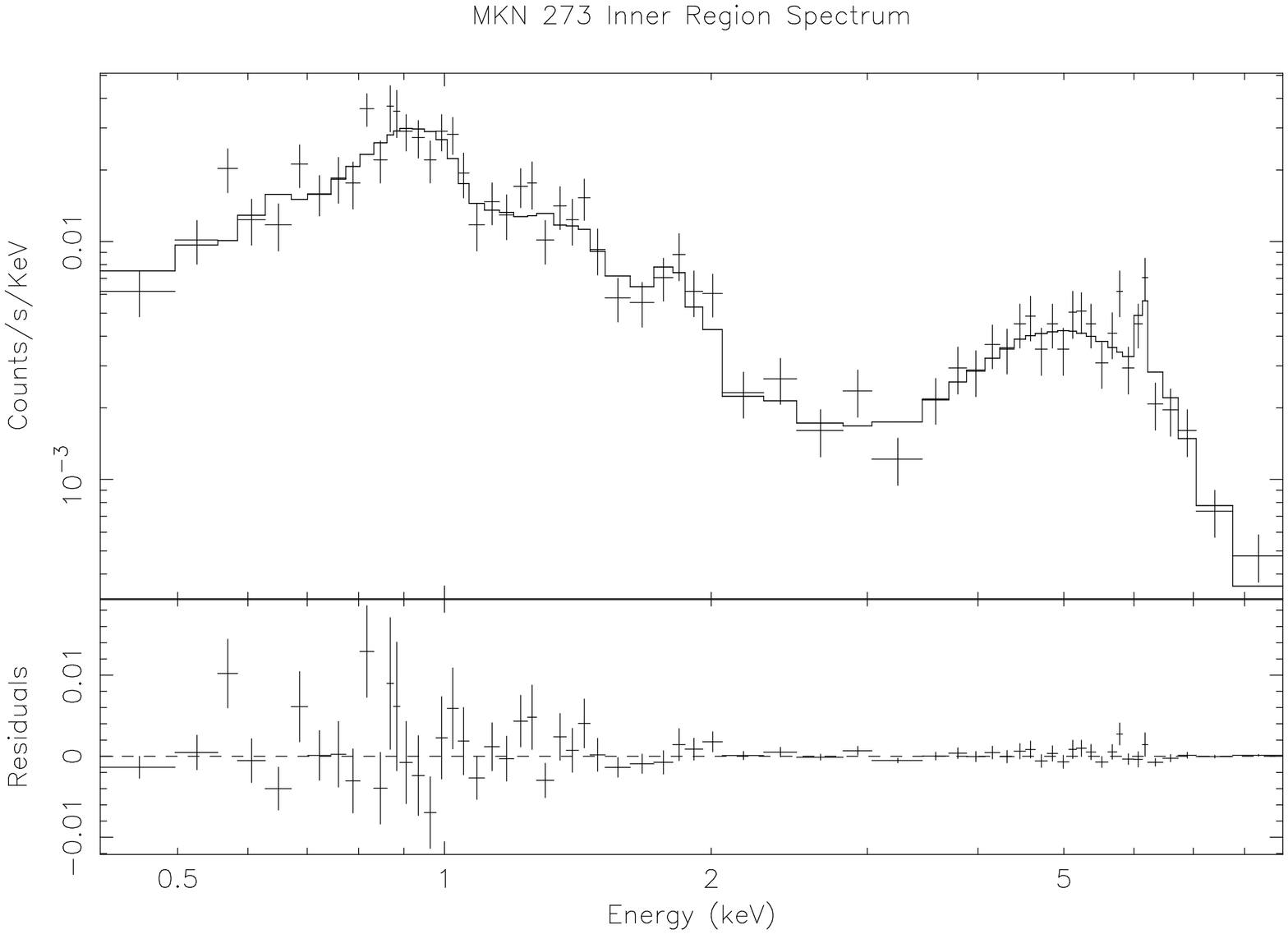}
\hfil
\includegraphics[width=1.8in,angle=-90]{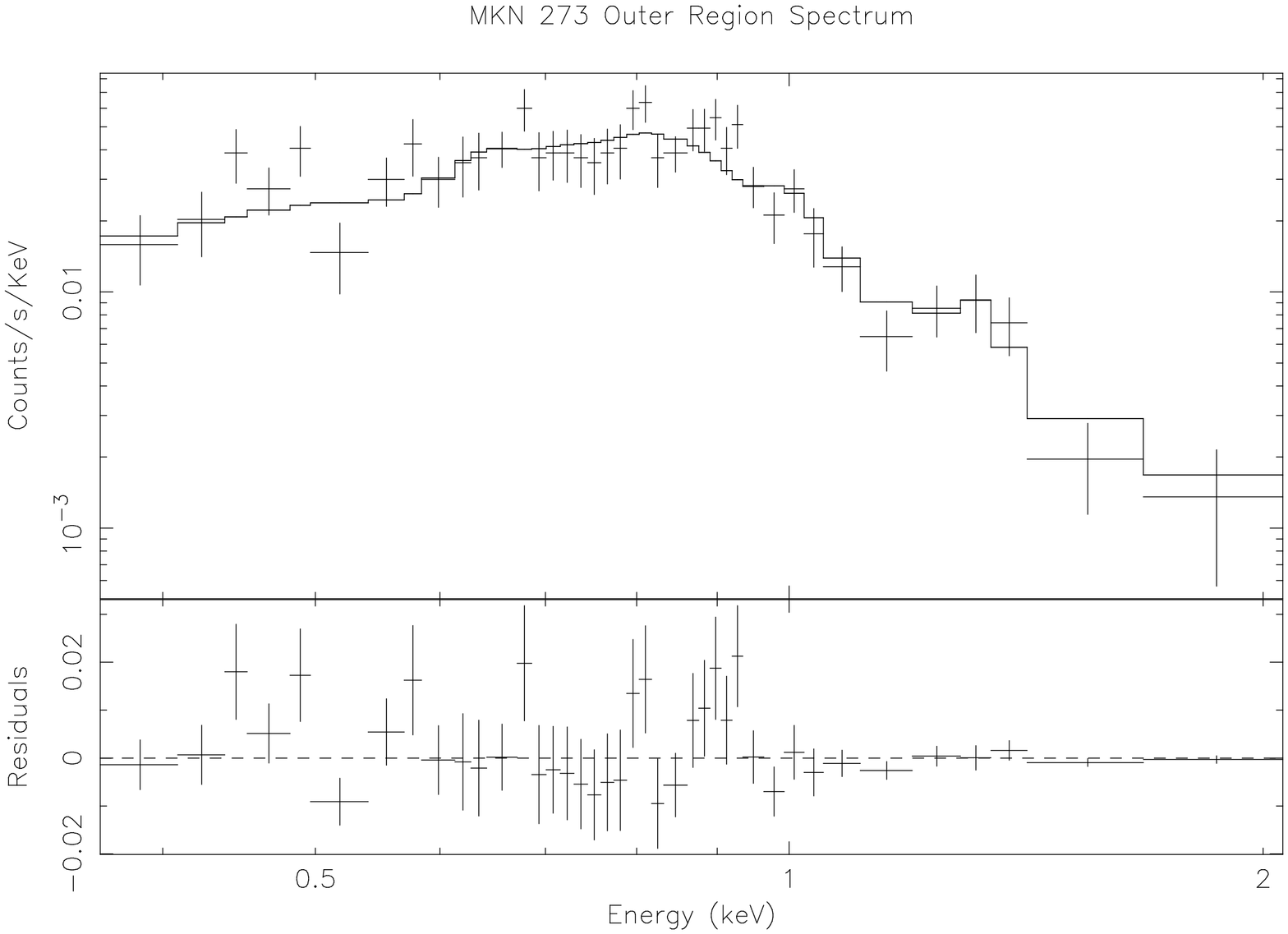}
\hfil
\vspace{.00in}
\includegraphics[width=1.8in,angle=-90]{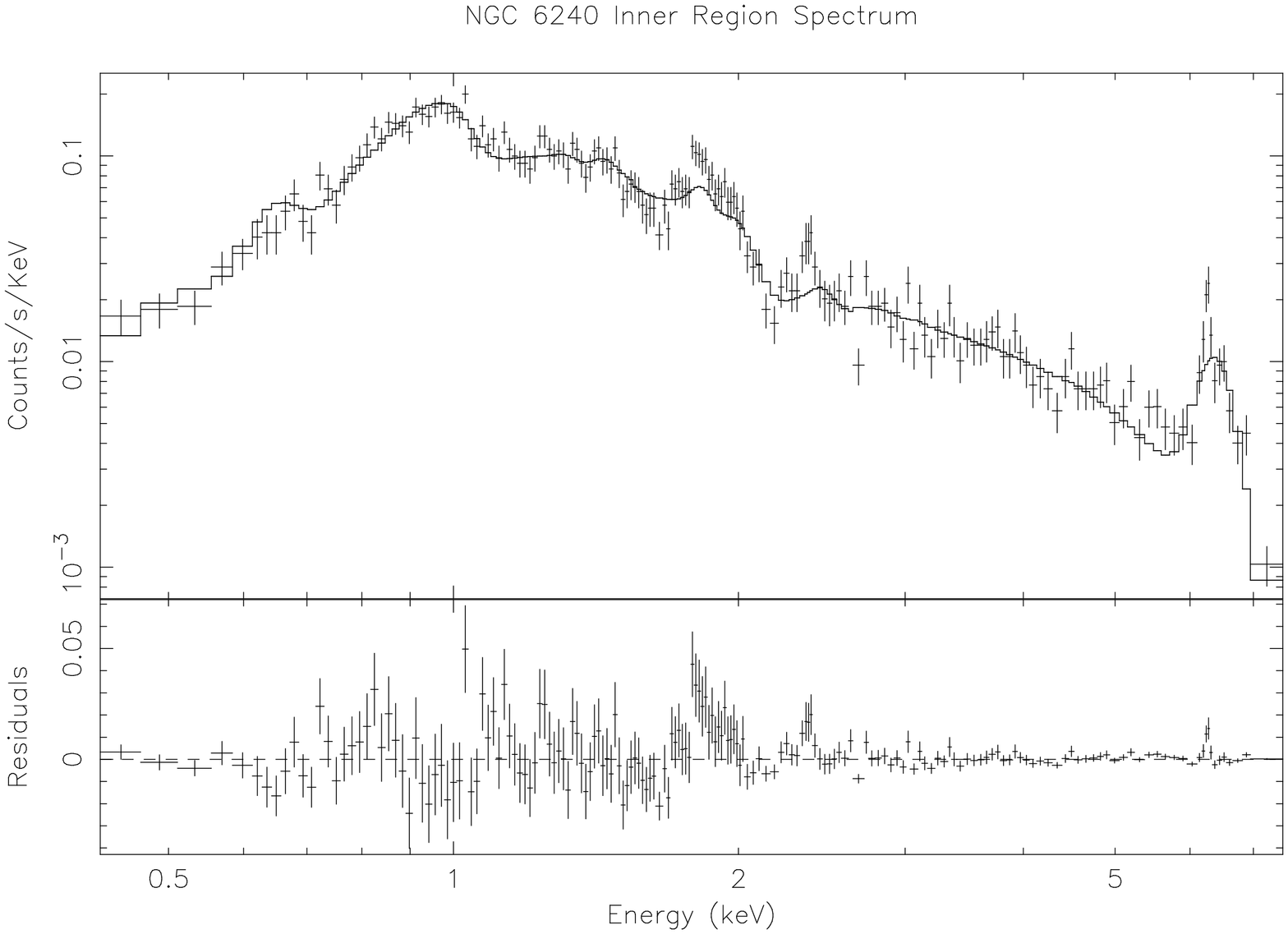}
\hfil
\includegraphics[width=1.8in,angle=-90]{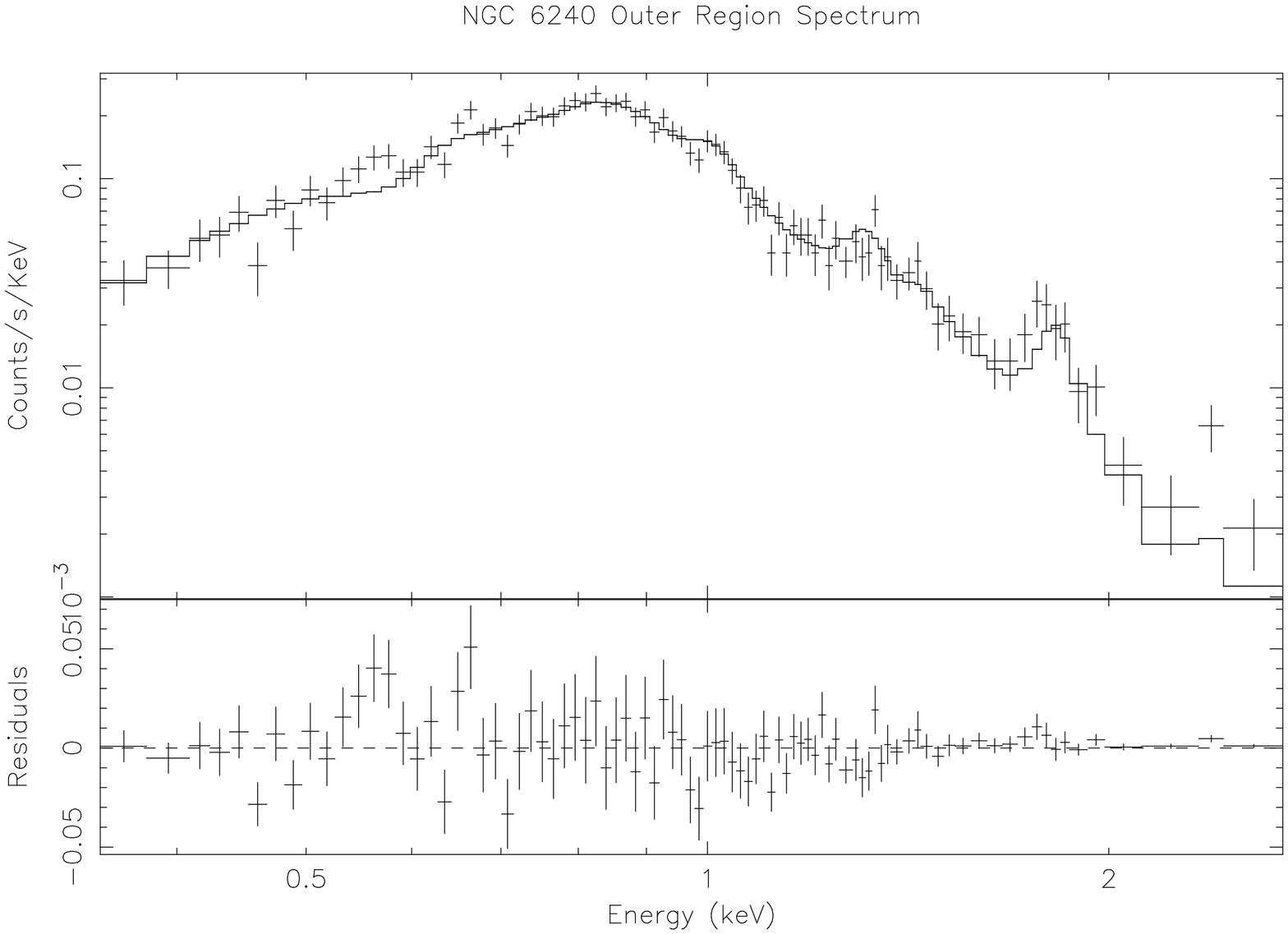}
\hfil
\vspace{.00in}
\caption{\ulg~Spectra: Spectra and Fits for the Inner and Outer Regions
for the \ulg~classified optically as AGN (Seyferts or LINERS) by
\citet{ptak03}.
\label{agnulgspectra}}
\end{figure}

\begin{figure}
\centering
\leavevmode
\columnwidth=.30\columnwidth
\includegraphics[width=1.8in,angle=-90]{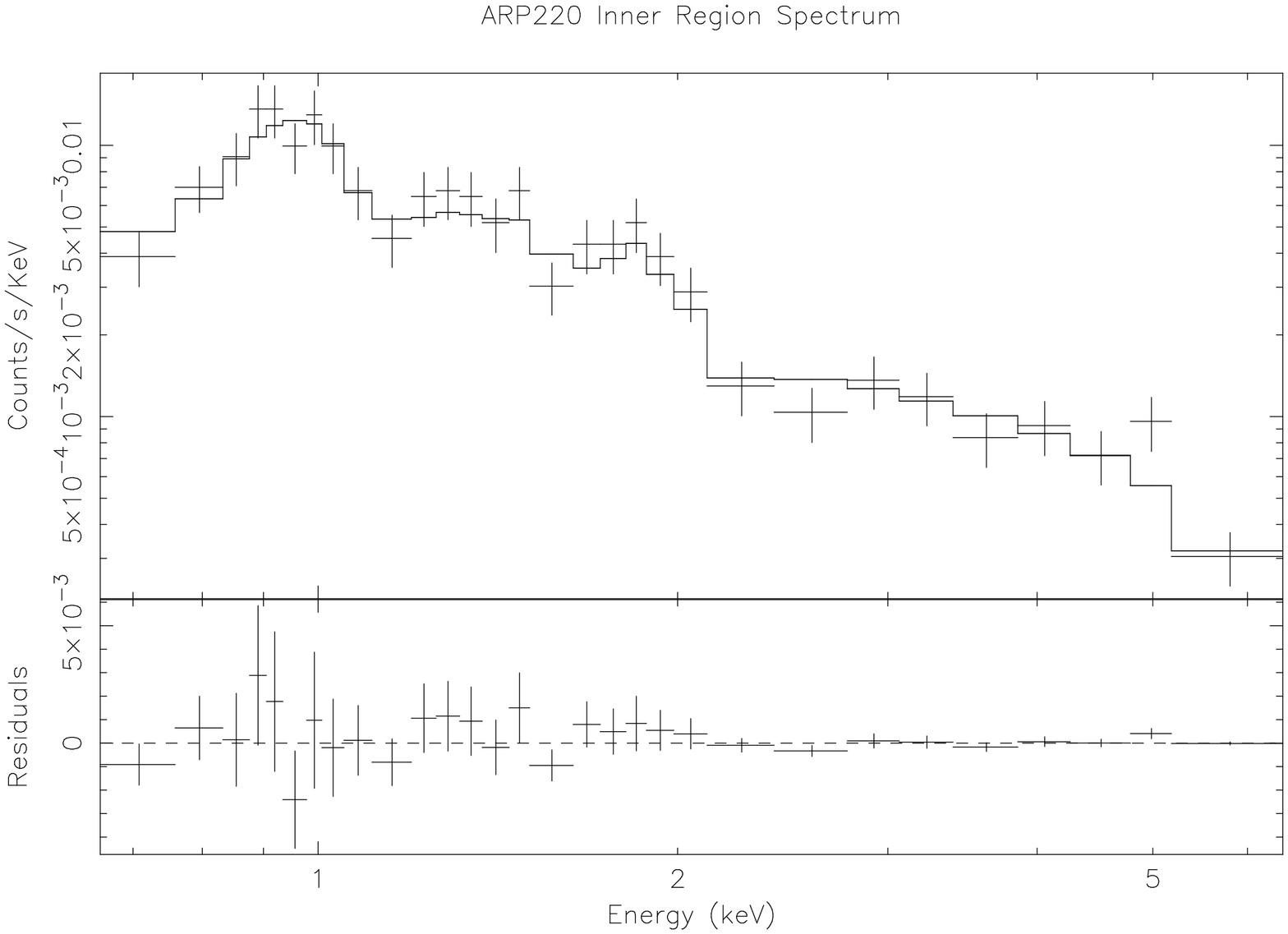}
\hfil
\includegraphics[width=1.8in,angle=-90]{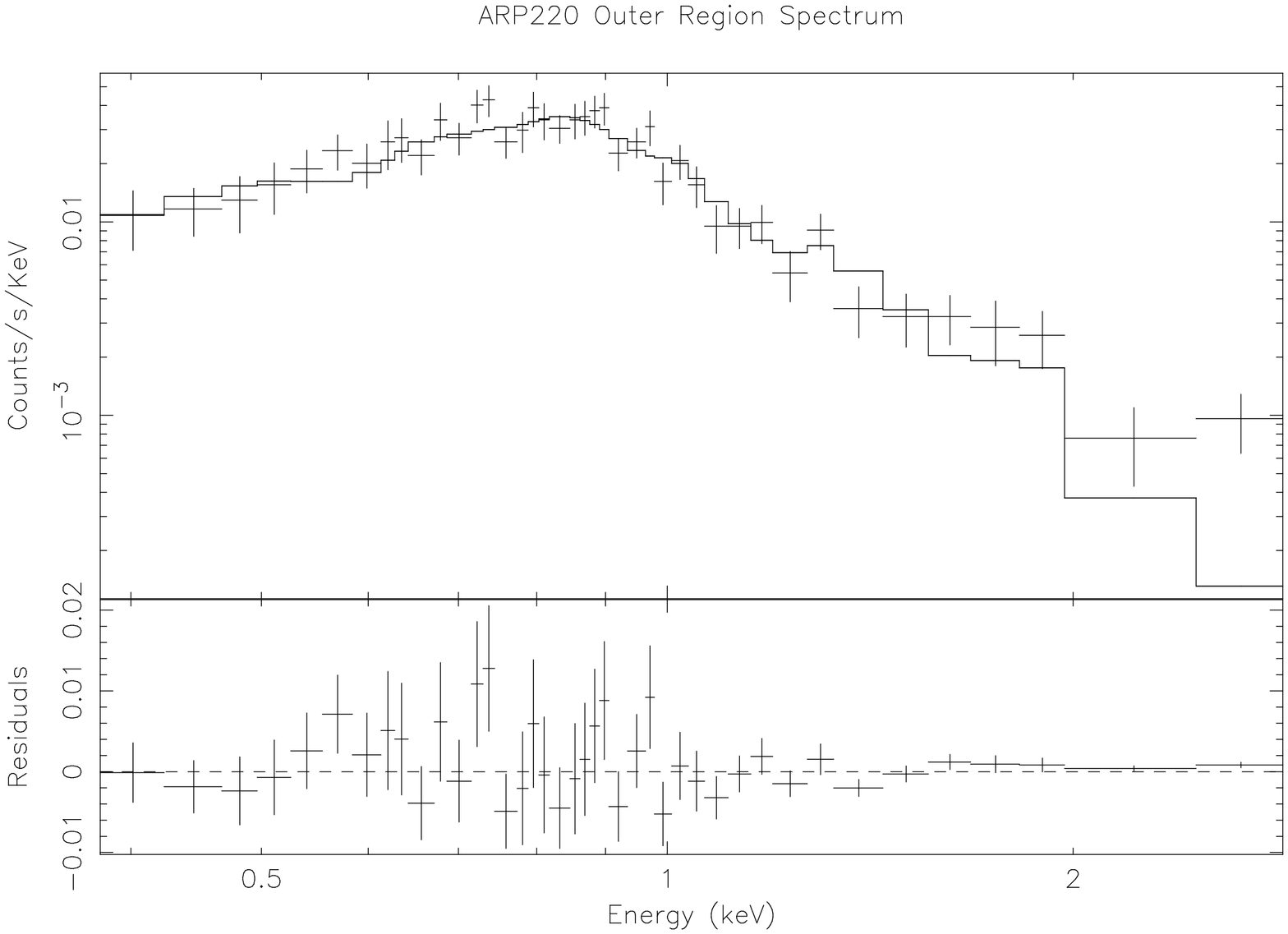}
\hfil
\vspace{.00in}
\includegraphics[width=1.8in,angle=-90]{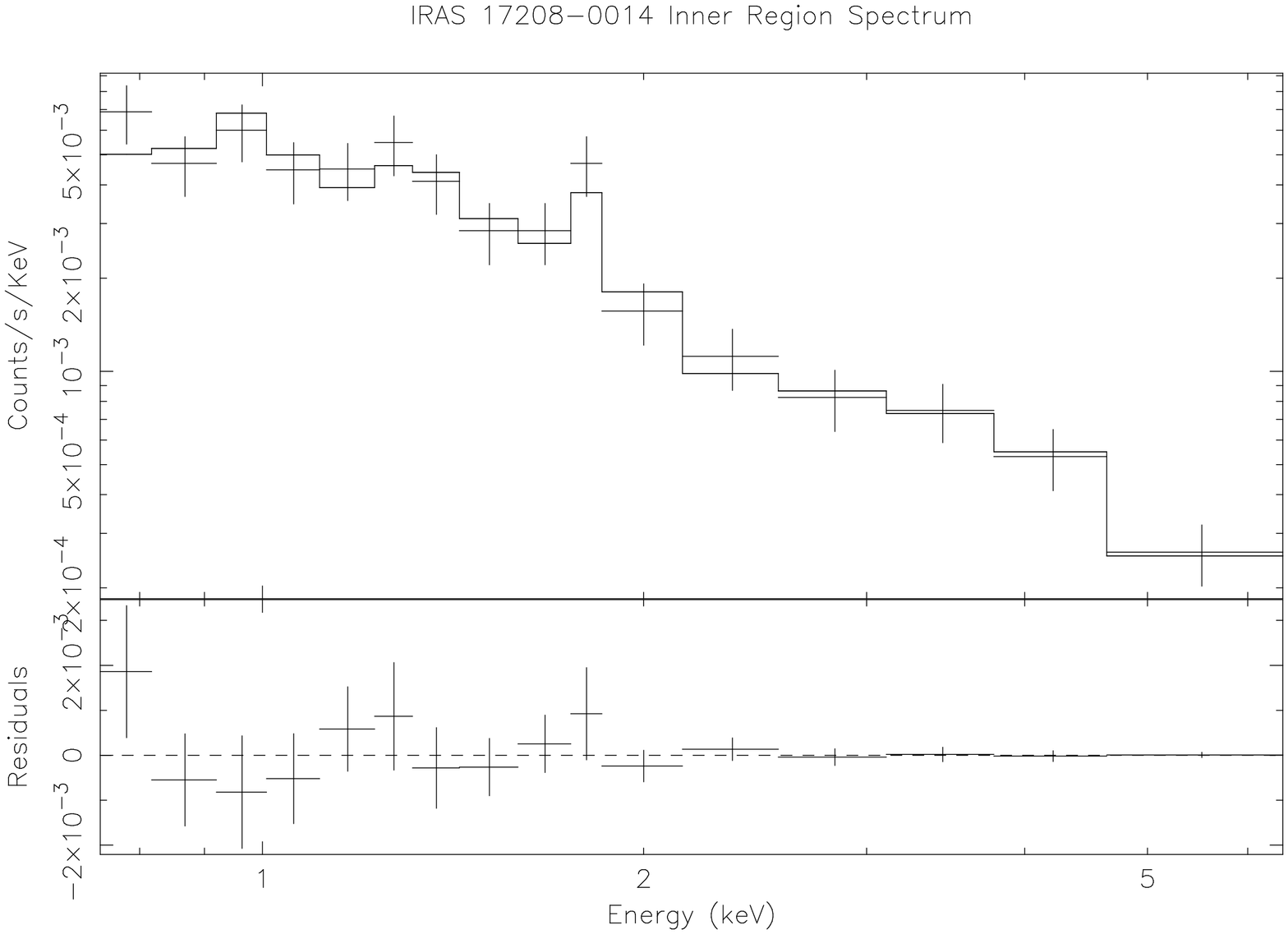}
\hfil
\includegraphics[width=1.8in,angle=-90]{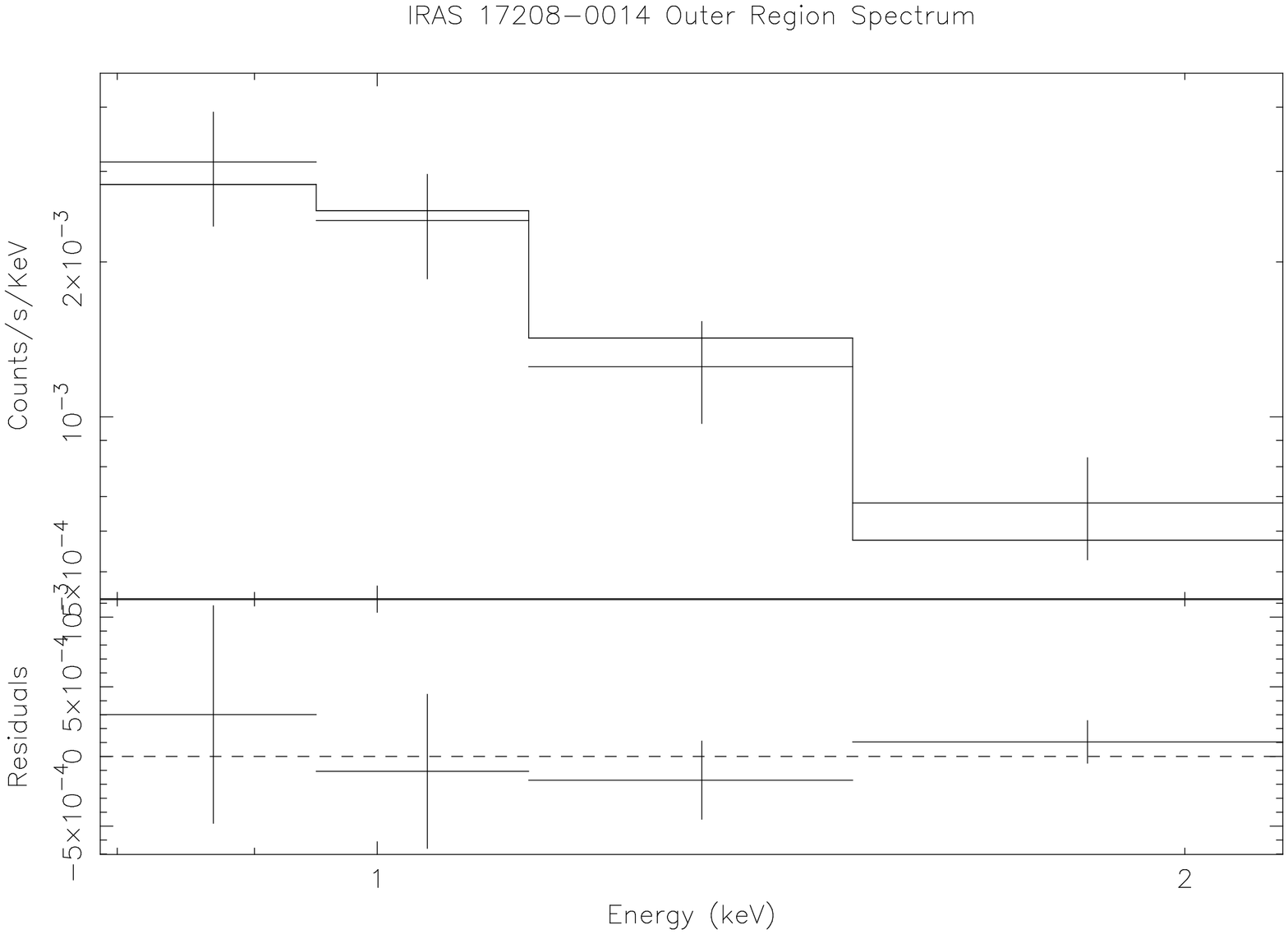}
\hfil
\vspace{.00in}
\includegraphics[width=1.8in,angle=-90]{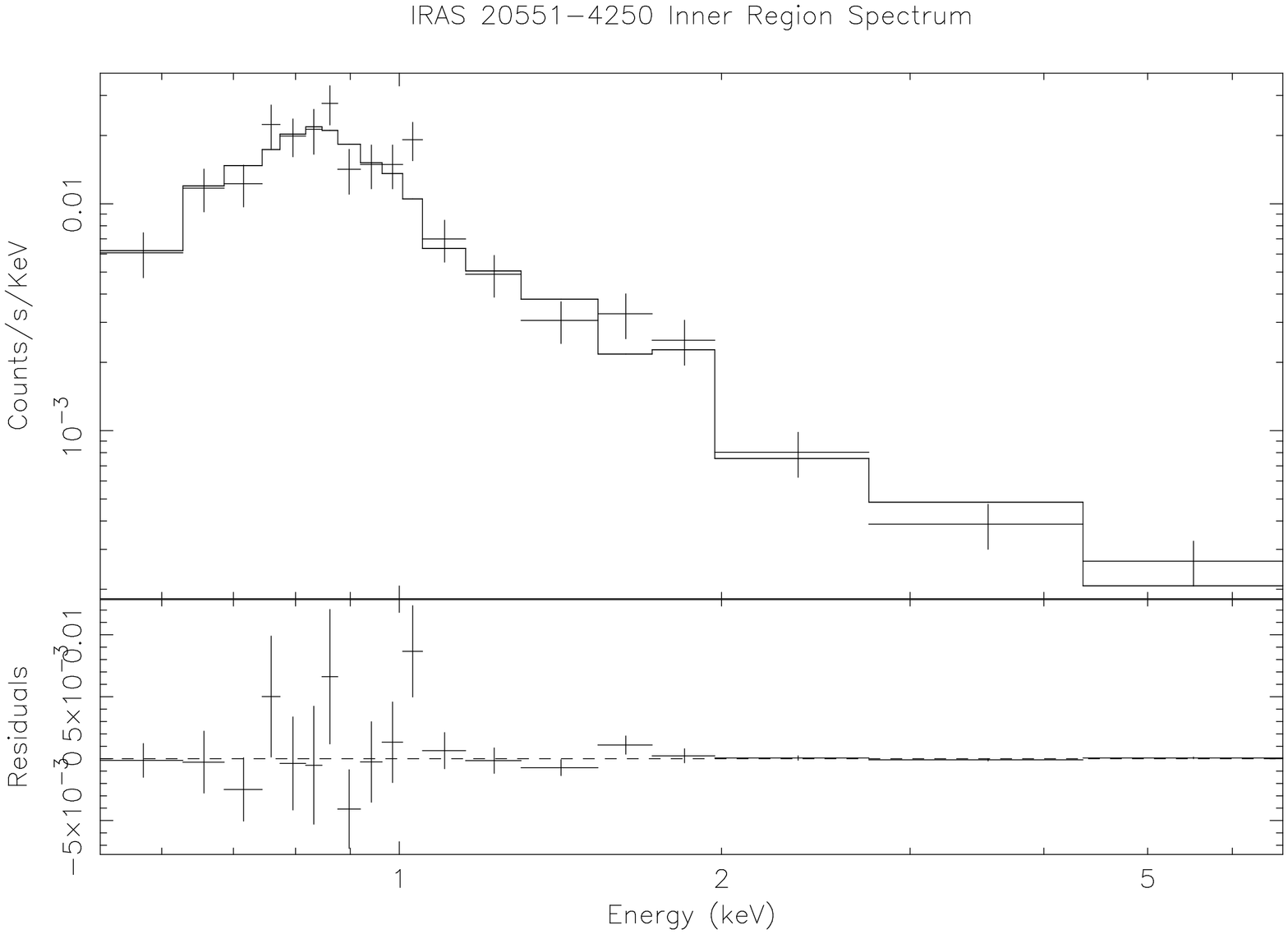}
\hfil
\includegraphics[width=1.8in,angle=-90]{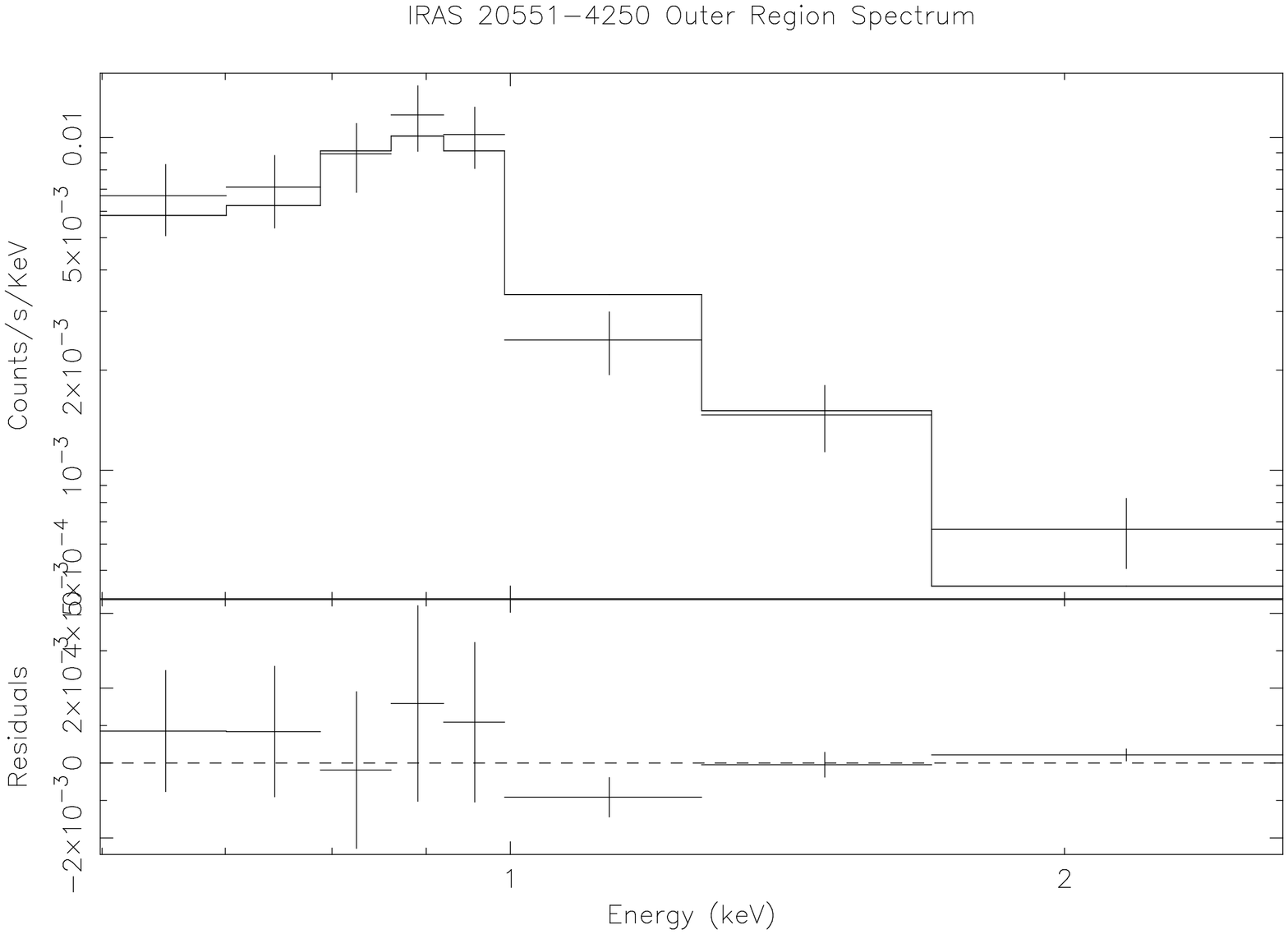}
\hfil
\vspace{.00in}
\includegraphics[width=1.8in,angle=-90]{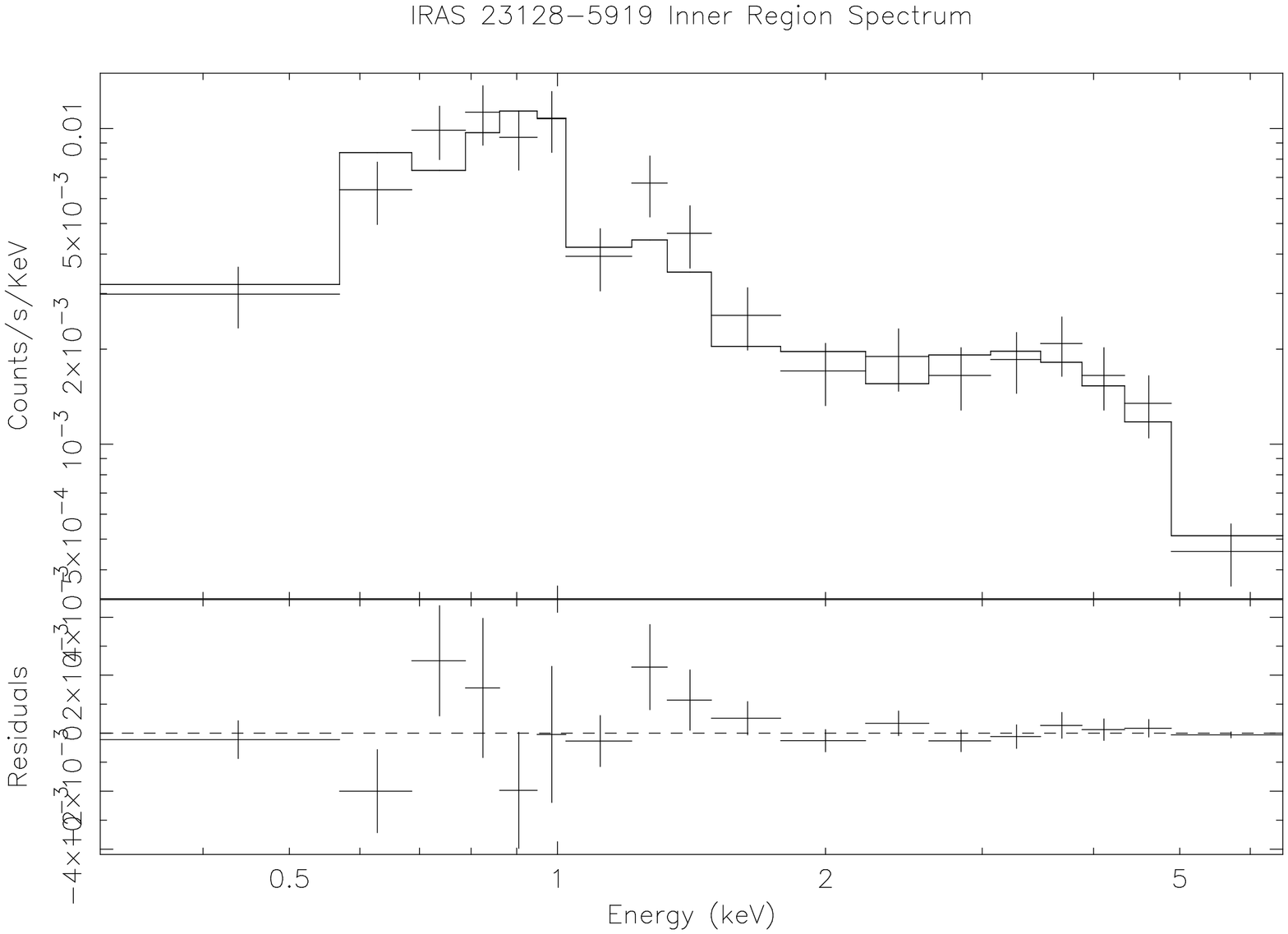}
\hfil
\includegraphics[width=1.8in,angle=-90]{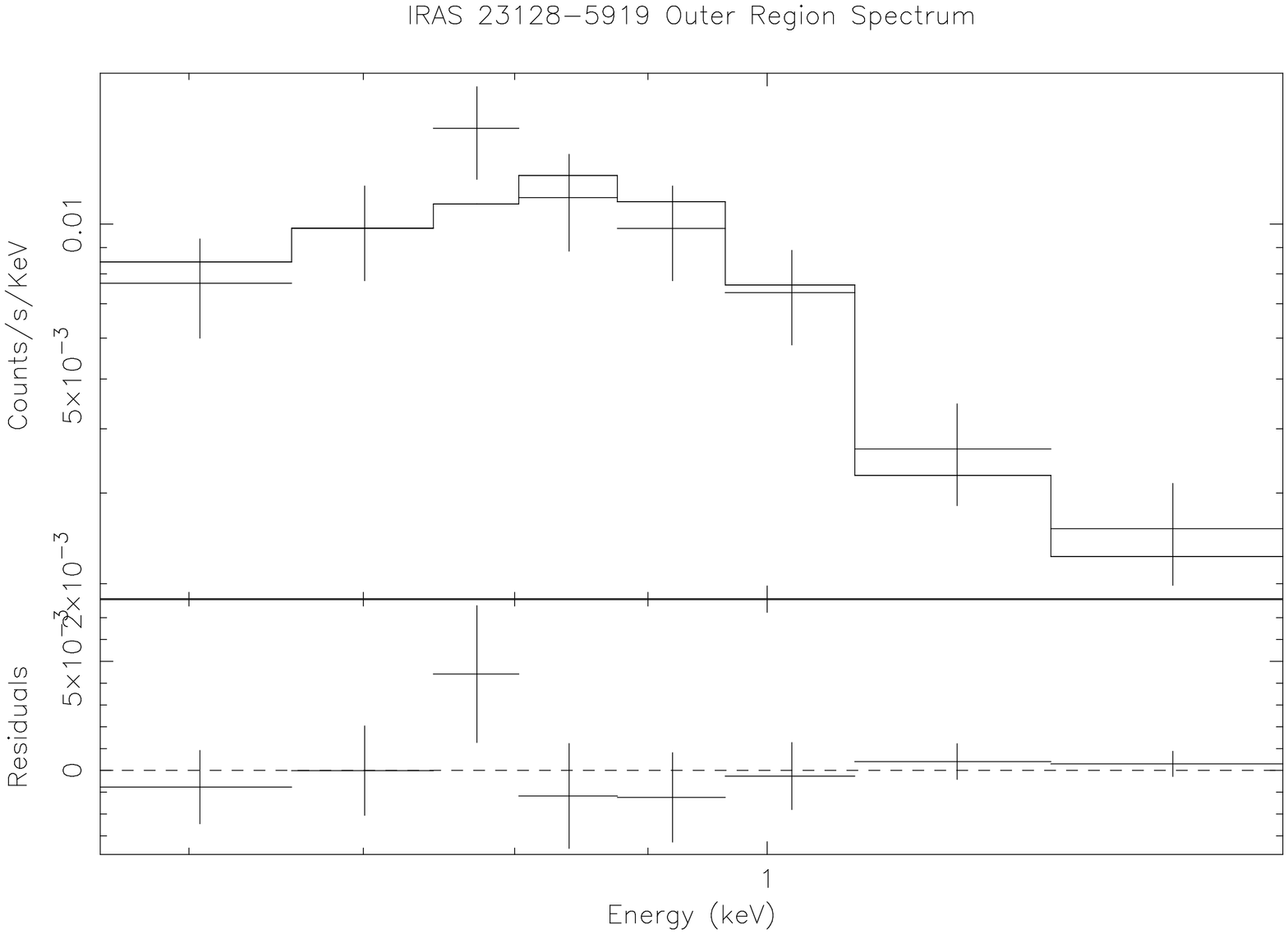}
\hfil
\vspace{.00in}
\includegraphics[width=1.8in,angle=-90]{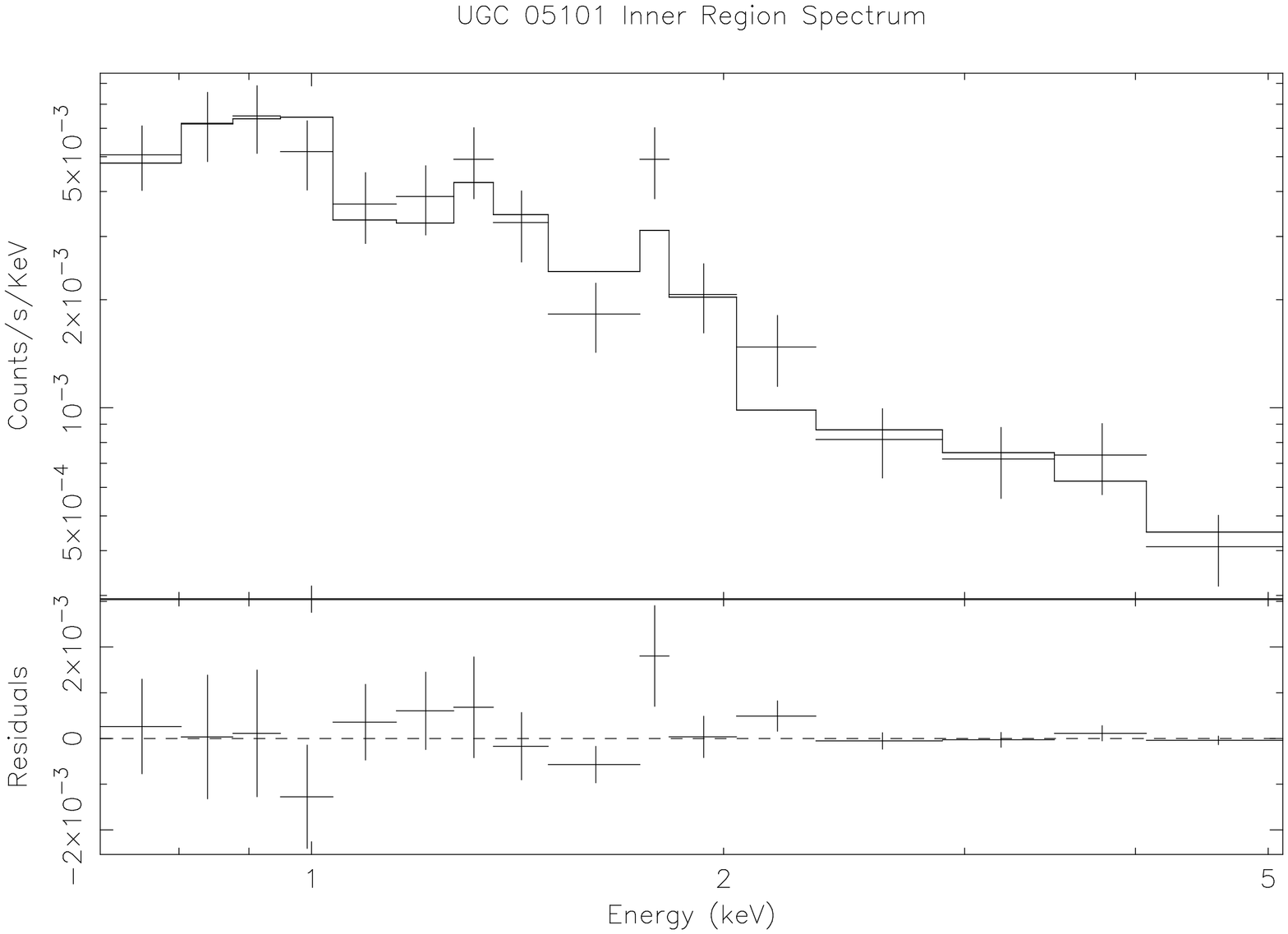}
\hfil
\includegraphics[width=1.8in,angle=-90]{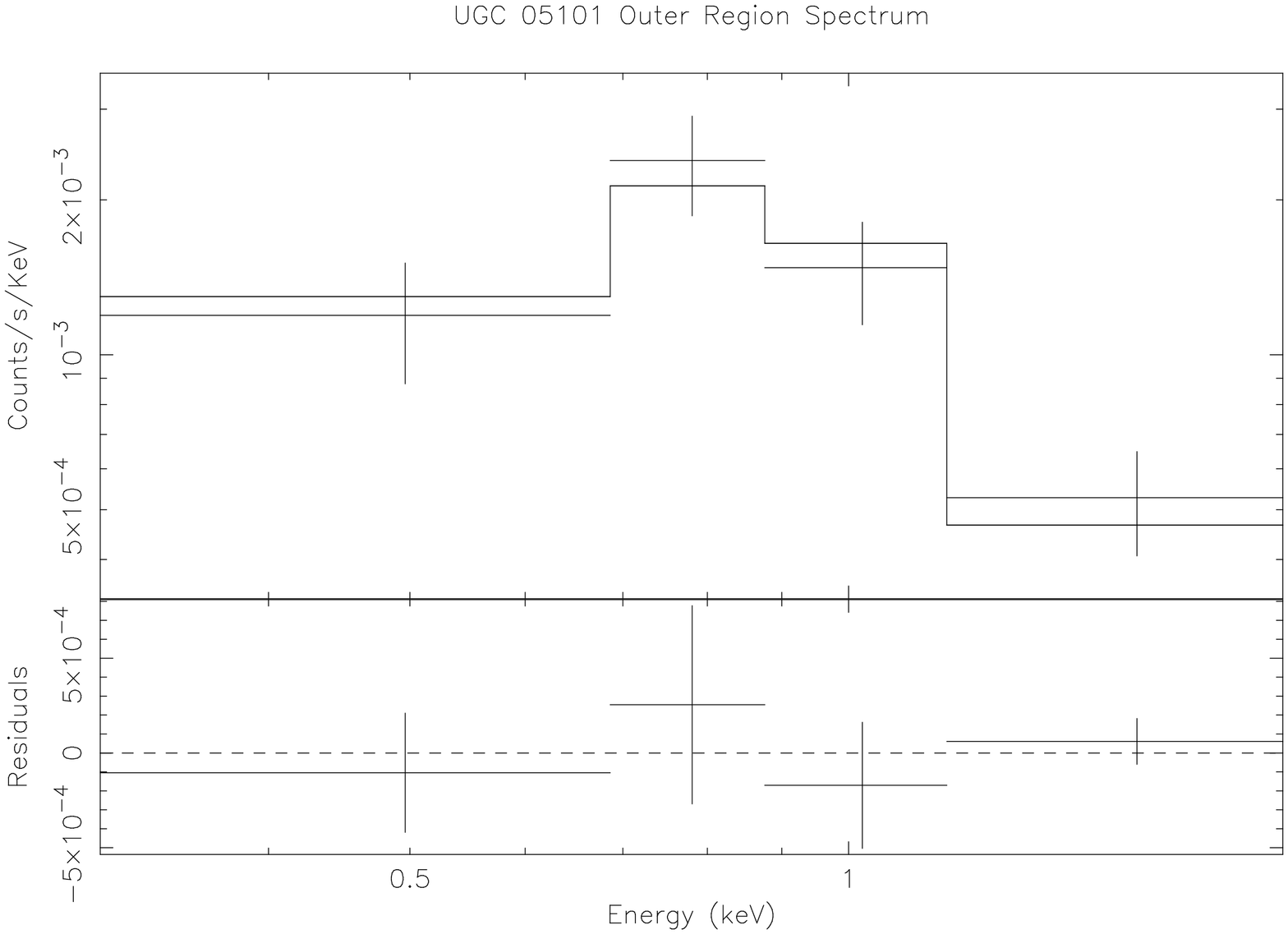}
\hfil
\vspace{.00in}
\caption{\scriptsize{\ulg~Spectra: Spectra and Model Fits for the Inner and Outer
Regions for the remaining \ulg.
\label{ulgspectra}}}
\end{figure}

\begin{figure}
\centering
\leavevmode
\columnwidth=.30\columnwidth
\includegraphics[width=1.8in,angle=-90]{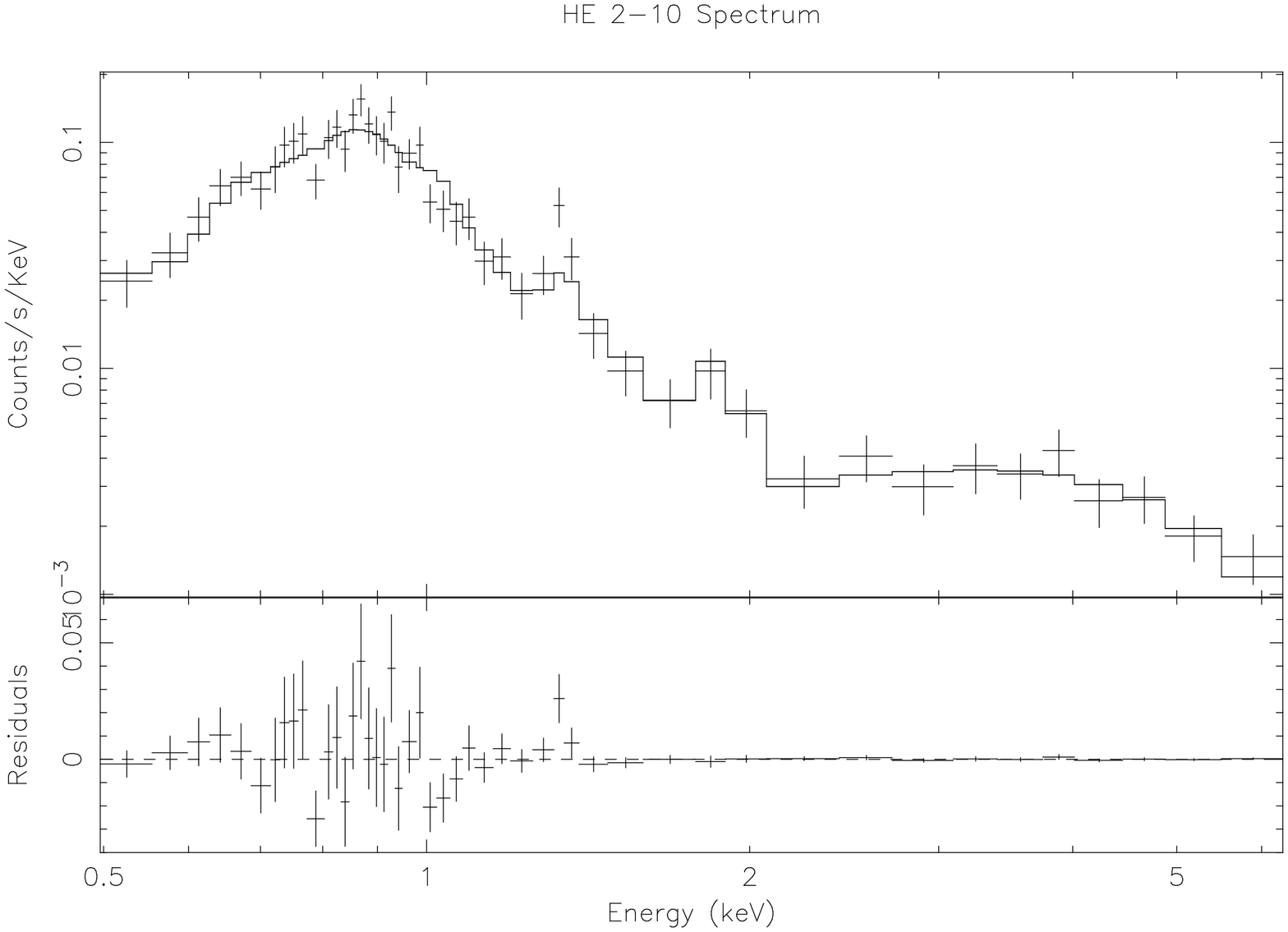}
\hfil
\includegraphics[width=1.8in,angle=-90]{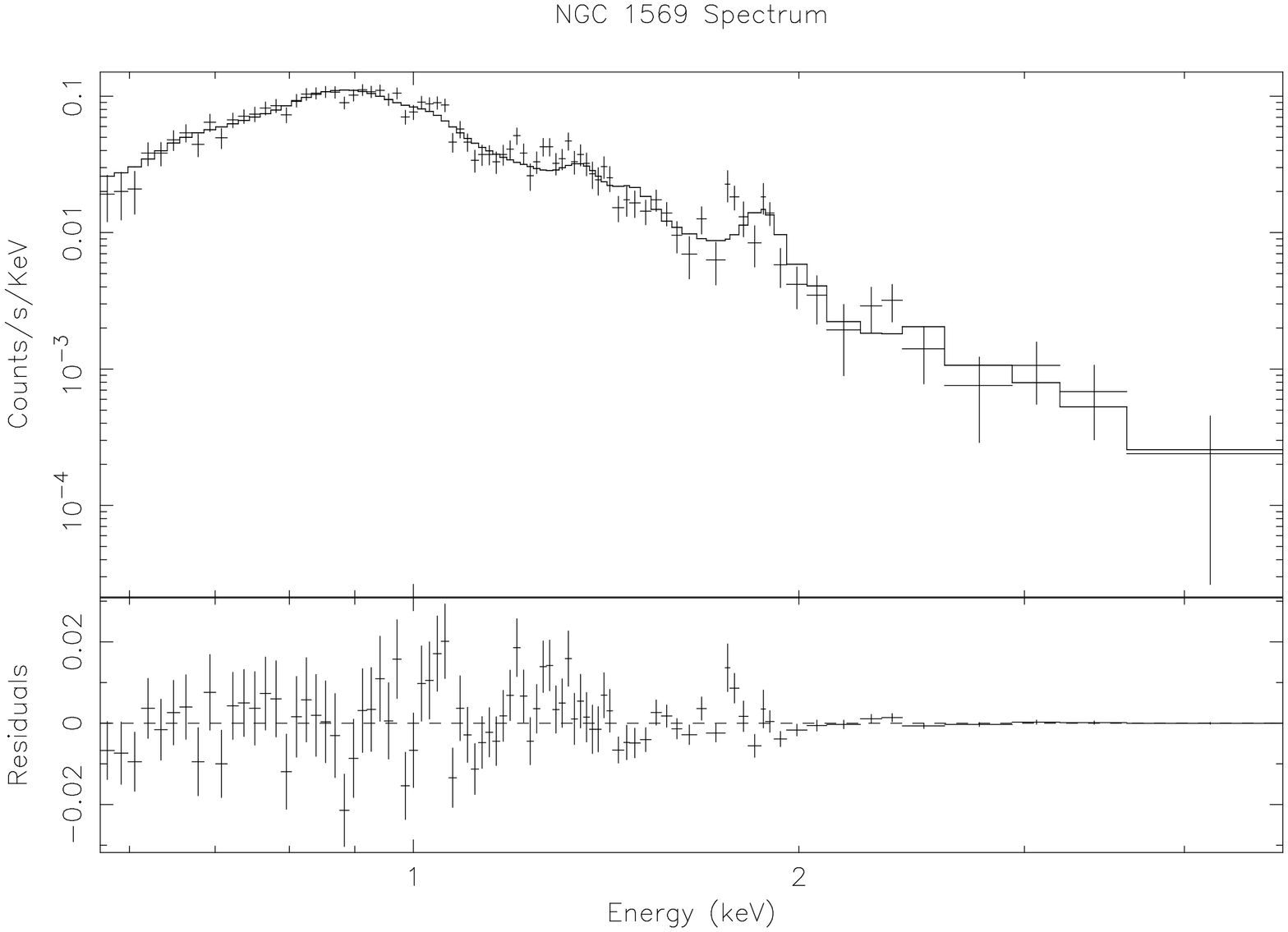}
\hfil
\vspace{.00in}
\includegraphics[width=1.8in,angle=-90]{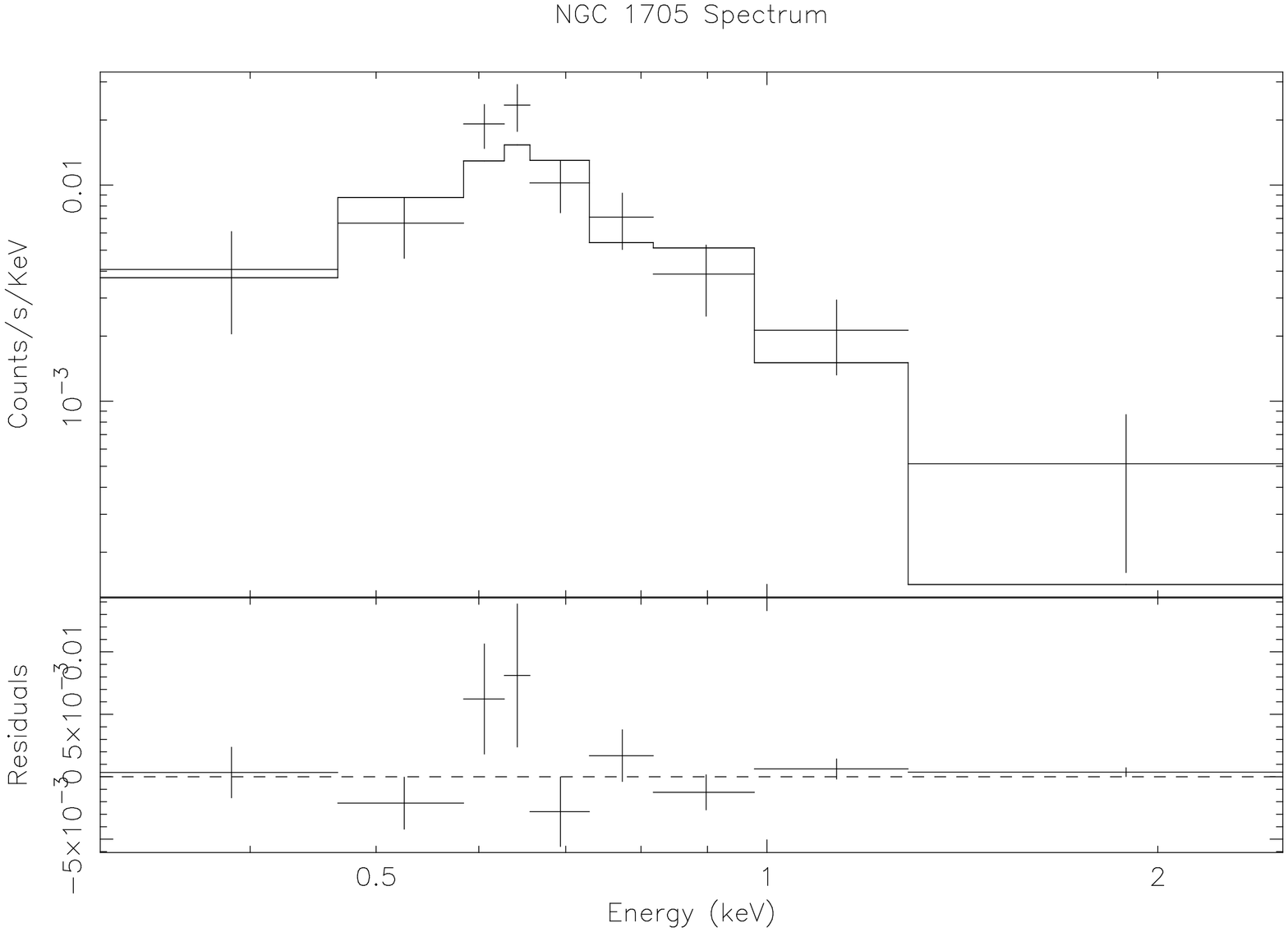}
\hfil
\includegraphics[width=1.8in,angle=-90]{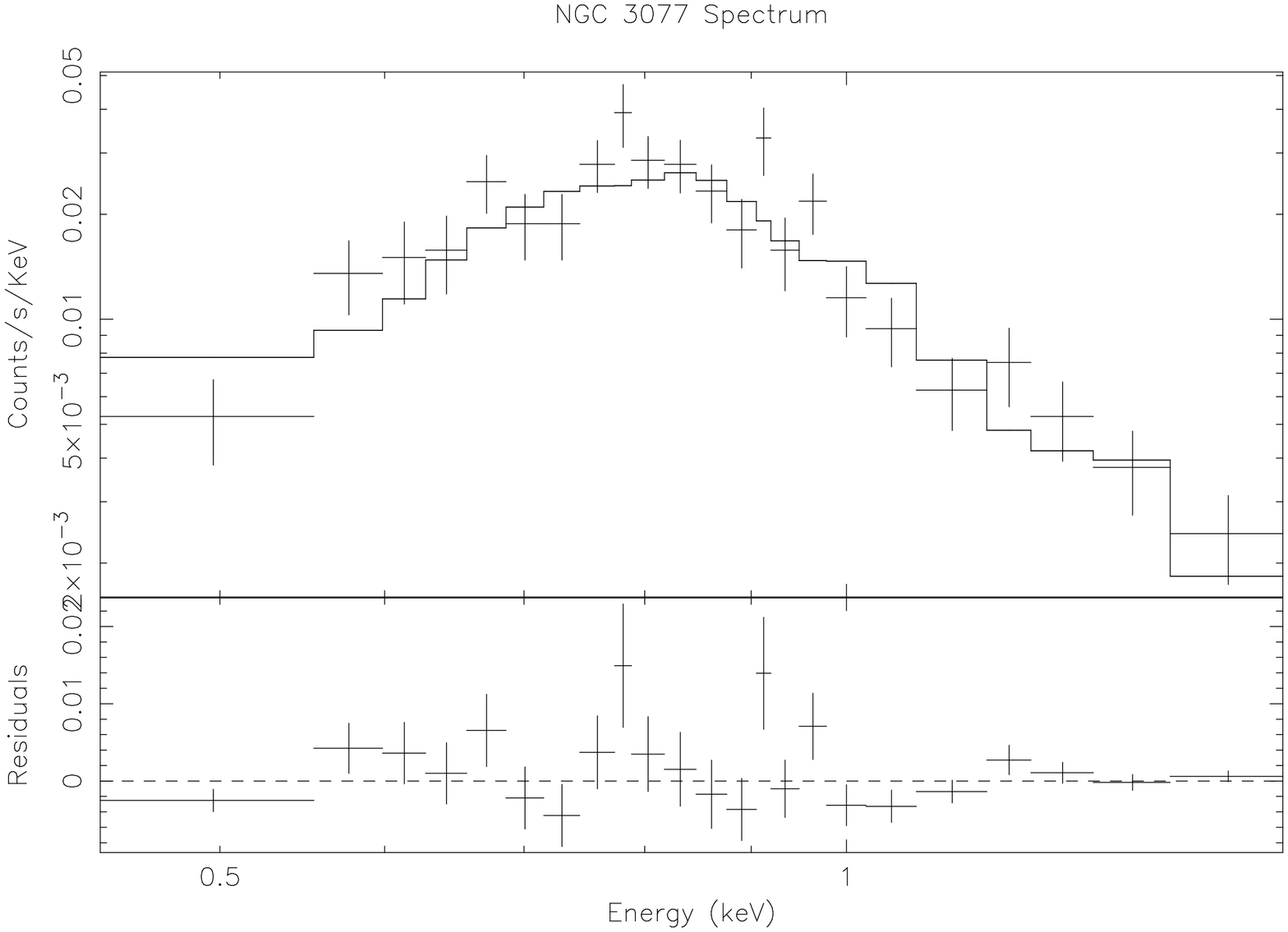}
\hfil
\vspace{.00in}
\includegraphics[width=1.8in,angle=-90]{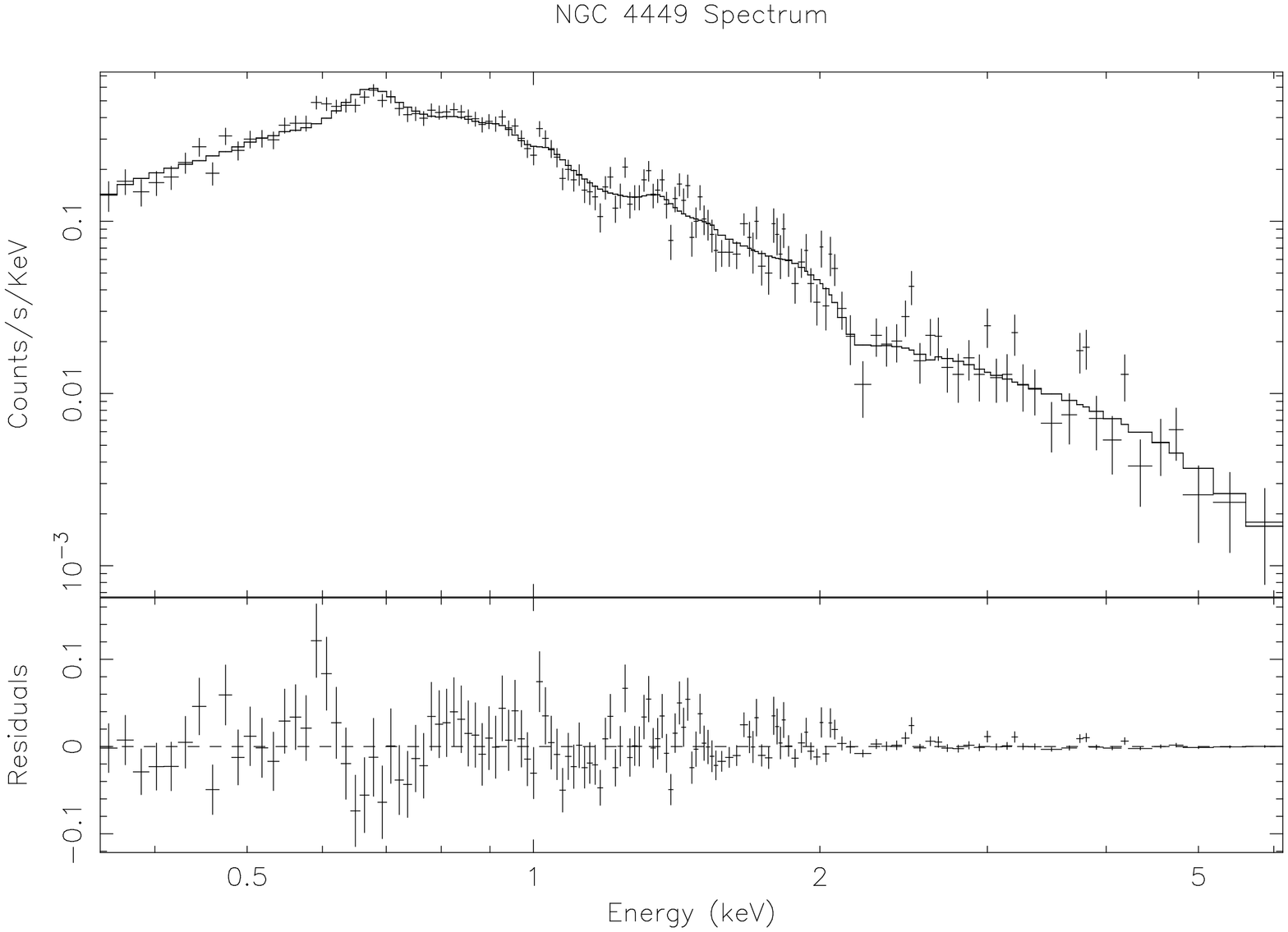}
\hfil
\includegraphics[width=1.8in,angle=-90]{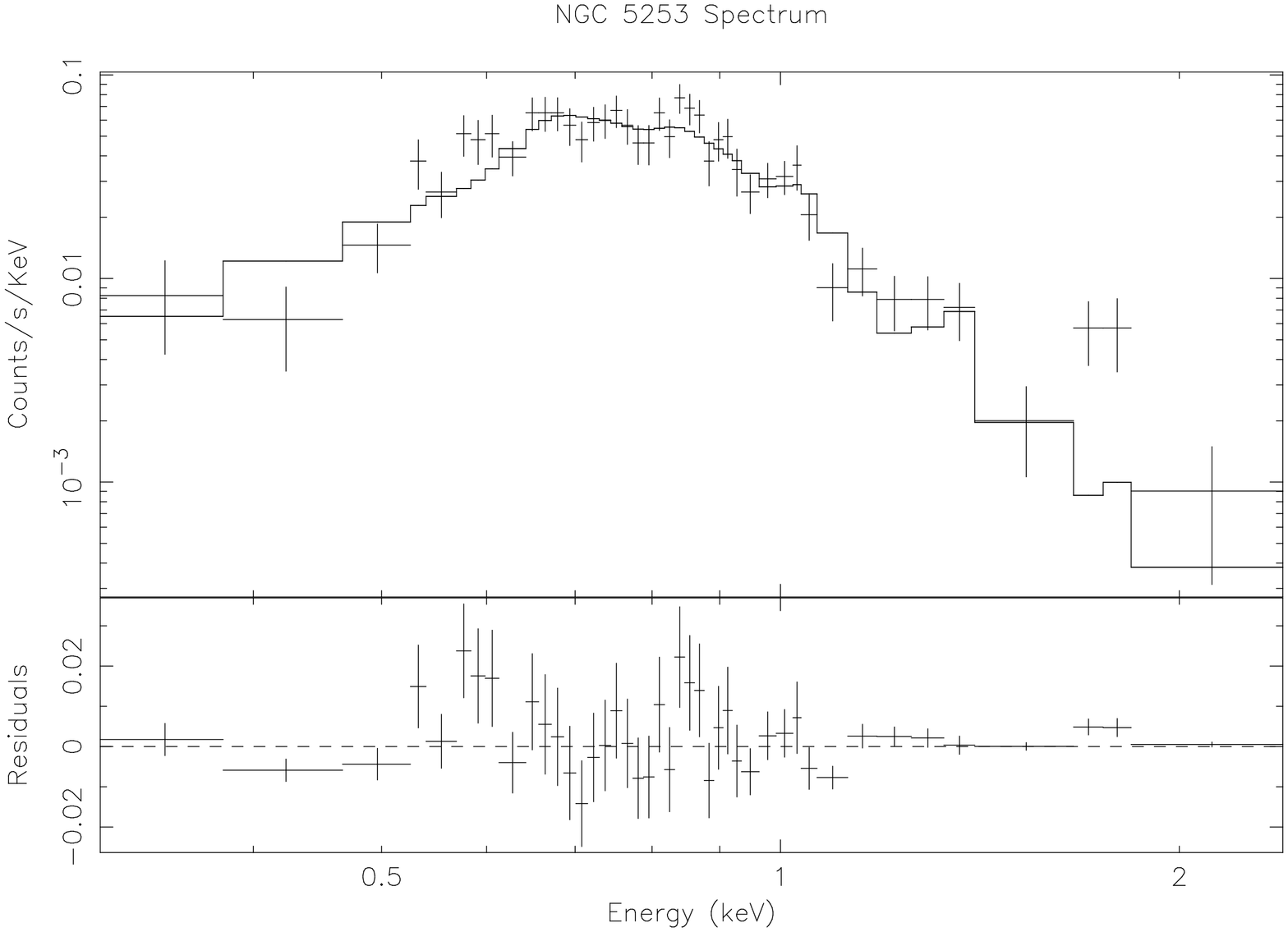}
\hfil
\vspace{.00in}
\caption{Dwarf Spectra: Spectra and Model Fits for the Dwarfs.
\label{dwarfspectra}}
\end{figure}

The extracted spectra are plotted in Figures 
\ref{agnulgspectra}, \ref{ulgspectra}, and \ref{dwarfspectra}.
A thermal component is clearly present in
all but the lowest quality data.  
As our objective is to compare the
global properties of our galaxies in a consistent manner, we have used a 
single thermal model in all of our fits (vmekal in xspec).
Although there is strong evidence for 
spatial variations in temperature for
several of our galaxies we have found generally, that a single
thermal model adequately represents each galaxy's
flux averaged properties.  While other spectral
models were examined, including two temperature models,
intrinsically absorbed temperature models, 
and simpler thermal models, these
models generally resulted in higher values of the
reduced \chsq.  As any thermal model would be a
simplification of the true multi-temperature
non-ionizational equilibrium physical conditions, 
we have used the simplest acceptable spectral fits
throughout our analysis.  This allows us to
analyze all of the galaxies in as uniform manner
as possible.  This consistency is important when
we compare the global properties of the X-ray gas
throughout our sample.

The vmekal model allows for
variations in elemental abundances.
However low signal to noise data and
weak or  unresolved emission lines make
it impossible to constrain many
of the elemental abundances.  Following
\citet{strick04a} we have therefore
tied many of the weakly constrained
abundances to other similarly evolving elements.
Mg, Si, Ne, S, and Ar are then tied to O
to form our \alphe~abundance.
We also tie the Fe abundance to Ni, Ca, Al,
and Na.  Ca has been included with the Fe elements
as it is similarly depleted onto dust grains and
has a negligible contribution to the X-ray emission
at these temperatures.  Although tied to other
elements in the vmekal model, the Fe and $\alpha$ abundances
are dominated by the contributions from Fe and O respectively.
The absolute $\alpha$ and Fe abundances are degenerate and relatively
unconstrained for many of the galaxies.
As we are primarily interested in \alphe~enrichment relative
to Fe we have only listed the  \afe~ratio for
which we can derive limits.

The spectra from the outer regions of the
\ulg~and most of the the dwarf starbursts 
have significantly less high energy
($>3.0~\rm{keV}$) emission than the inner 
regions of the \ulg.
For several there is no high energy
emission detected.  For this sample
we have restricted our
fits to only include the 0.3 - 3.0 keV
energy band.  A single thermal plasma (vmekal)
model, without an additional powerlaw component,
provides an excellent fit for the 8 \ulg~with
detectable outer region emission and 5 of the
dwarf starbursts.

Fitting models to the outer regions of the \ulg~is complicated by
low number counts in several of the \ulg.  In the case
of \irof~there are too few counts (if any) to attempt
spectral fitting.  The other eight \ulg~do have
detectable diffuse emission ranging from 76 counts
above background in \ugc~to over 4100 counts for
NGC~6240.

An absorbed powerlaw was required to fit the inner regions
of the \ulg~and the dwarf galaxies HE 2-10 and NGC 4449.   
The redshift of the absorption column was set
to that of the host \ulg~and the column density was
allowed to vary freely.  As the
flux above 3.0 keV is dominated by the
powerlaw emission we have fit these spectra
in the 0.35 - 8.0 keV energy region.  This allows 
a much more accurate determination of the
powerlaw parameters.

In \citet{ptak03} \feka~emission was strongly detected in NGC~6240
and MKN~273.  For the inner region in NGC~6240
we have added a gaussian component centered at 6.4 keV.  This
provides an excellent fit for the spectrum of NGC~6240.
The fit to the inner region of MKN~273 however
is more complicated.  To fit the
\feka ~line we have also added a gaussian.  However we have
found that an additional unabsorbed powerlaw is required.
We have set the unabsorbed powerlaw to have the same slope
as the absorbed powerlaw.  The additional powerlaw is motivated
by examining representative color images of MKN~273.  The
central source is strongly absorbed along
some lines of sight but not others.  The final fit
to MKN~273 gives us an acceptable reduced \chsq$=1.2$.

The inner spectra of \irto~suggests the need for a multiple
temperature model.
Although a two temperature model improves the reduced \chsq~from 1.6 to
1.4 the derived parameters are not significantly affected.  The derived  
flux weighted temperature from the two temperature model is consistent with
the single temperature model.  The powerlaw, redshifted hydrogen
absorption column, \afe, and flux are also found to be similar for both  
fits.  For consistently with the other spectral fits we have used the results
from the one temperature model for \irto.

The set of spectral fitting parameters for both the
inner and outer regions for all the \ulg~ and for
the dwarf starbursts can be found in Table \ref{spectra}.

\subsection{K-Band, FIR and UV Luminosities}
\label{sec:other_luminosities}

To characterize the basic properties of the starbursts and their host
galaxies, we have assembled a combination of near-infrared (K-band),
far-infrared (IRAS), and vacuum ultraviolet data from the literature  
(see
Table \ref{lumin}).

We will use the near-IR luminosity as a rough tracer of the stellar
mass of the galaxy, since the K-band mass-to-light ratio is
only a weak function of the age of a stellar population  
\citep[e.g.][]{bell01,bell03}.
We note however that this approximation
may not be valid for all the ULIRGs, in which the extremely high
star formation rate and possible presence of an AGN mean that
the K-band light will not necessarily be dominated by the older stars
that dominate the galaxy mass \citep[e.g.][]{genz01}.  In fact, the AGN
is known to dominate the K-Band luminosity of MKN 231, so we
have dropped MKN 231 from our results involving its K-Band luminosity.
The K-Band Cousins-Glass-Johnson luminosities were derived
for every galaxy in our sample
using the 2MASS Large Galaxy Atlas \citep{jarr03}.  The K and J Band
fluxes given in the catalog were converted to K-Band CGJ using the
transformations from \citet{carp01}.

The bulk of the bolometric luminosity of the ULIRGs and edge-on
starbursts
emerges in the far-infrared. \citet{ptak03} argue that the hard
X-ray properties of the ULIRGs (with the exception of the near-ULIRG
NGC~6240) do not support the idea that an AGN makes a dominant
contribution to the far-infrared luminosity. We will therefore use
the far-infrared luminosity as a proxy for the star formation rate
in the ULIRGs and edge-on starbursts. The dwarf starbursts are
substantially less dusty than the other two classes, and a significant
amount of the starburst luminosity escapes in the UV. Following
\citet{heck98}, we will use the sum of the UV and far-infrared
luminosities as a proxy for the star formation rate in the dwarf
starburst.

The {\it IRAS} fluxes \citep{soif89,rice88}
and transformations from \citet{sand96} were
used to calculate FIR luminosities.
Note that no IRAS color correction
term has been applied to our FIR luminosities.
The UV flux measurements we used have mean wavelengths in the range
$\lambda = 1500$ to $1900~\AA$. UV
fluxes are defined as $\lambda F_{\lambda}$.
The UV flux measurement for NGC 1569 comes from \citet{isra88},
for NGC 1705 from \citet{rifa95}, for NGC 3077 from \citet{code82},
for NGC 4214 and NGC 5253 from C.~Hoopes (private communication,
based on UIT observations),
and for NGC 4449 and He 2-10 from NED (based on OAO and IUE  
observations).
The UV fluxes were corrected for foreground Galactic extinction at the  
effective wavelength
of the appropriate filter
using the extinction law of \citet{card89} and the values of the
Galactic optical extinction from \citet{schl98}.
Luminosities are listed in Table~\ref{lumin}.

\section{Discussion}

In this section we will compare the principal properties of the
diffuse X-ray emission in our sample of dwarf starburst, edge-on
starbursts, and ULIRGs. We begin
by comparing the structure and morphology
of the diffuse X-ray emission and its morphological connection
to the warm optically-emitting plasma. We will then examine how
the size and luminosity of the source of diffuse emission
scales with the luminosity (the star formation rate) of the
starburst. Finally, we will use the spectra to examine the thermal
and chemical properties of the hot gas.

\subsection{Structure \& Morphology}

\begin{figure}
\includegraphics[width=6in,keepaspectratio=true,clip=true,origin=bl]{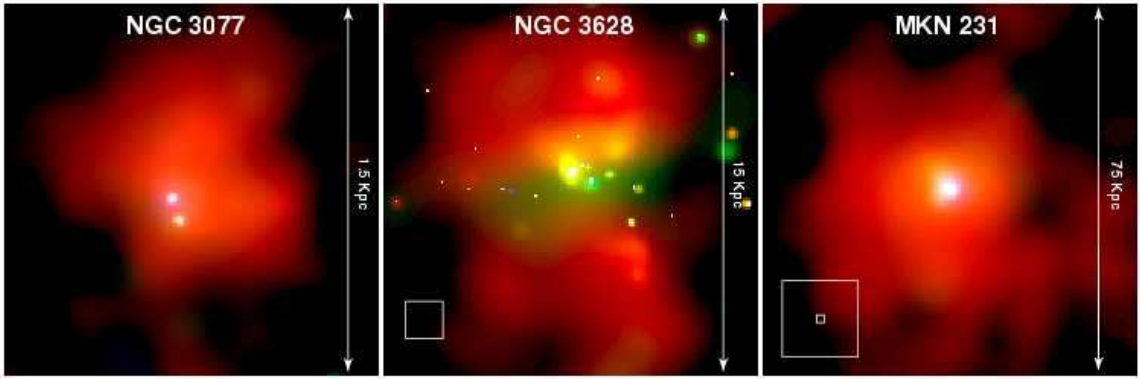}
\caption{Galaxy Comparison:  Adaptively Smoothed Representative Color X-ray Images
0.3-1 keV Red, 1-2 keV Green,
2-8 keV Blue;  From the left to the right we have the
dwarf starburst NGC 3077 ($\rm{1.5kpc\times 1.5kpc}$), the edge-on
starburst NGC 3628 ($\rm{15kpc\times 15kpc}$),
and an AGN \ul~MKN 231 ($\rm{75kpc\times 75kpc}$). The dimensions
in the parenthesis
are the physical size of the viewable area for each galaxy.  Also, for
comparison between the
panels, in the middle plot and right panels we have placed a box of size
$\rm{1.5kpc\times 1.5kpc}$ and in the panel on the right we have  
additionally
placed a box of size $\rm{15kpc\times 15kpc}$.
There are strong morphological similarities between the galaxy types
even though they span almost a factor of 50 in physical scales and  
nearly four
orders of magnitude in star formation rate.
\label{3types}}
\end{figure}

A simple visual comparison of the morphology
suggests a similar physical origin for the diffuse X-ray emission in
dwarf starbursts, starbursts, and ULIRGs
(Figure \ref{3types}).
These three representative color images are very similar to one another,
not only in morphology, but in ``color'' (the broad-band X-ray spectral
shape).
Although the physical scales differ
by almost a factor of 50 from dwarf to ULIRGS, it is difficult to
separate these galaxies based on the morphology of their diffuse
X-ray emission.  The biggest visible difference is in the
edge-on starburst galaxy NGC 3628,
but this is due to photoelectric absorption
from
the edge-on large scale ISM of the galaxy
(i.e. it is simply an orientation effect).

\begin{figure}
\centering
\includegraphics[width=3in,angle=-90]{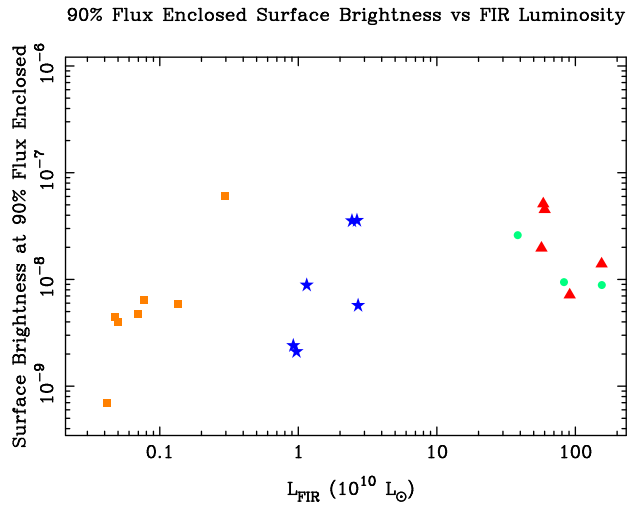}
\caption{
Mean Enclosed X-ray Surface Brightness  
($\rm{photons~s^{-1}cm^{-2}arcsec^{-2}}$
in the 0.3 to 1.0 keV band)
at the 90\% Enclosed Flux Radius  vs FIR  Luminosity
The X-ray surface brightness remains fairly constant
throughout the sample showing that the X-ray surface brightness is not
correlated with the SFR.
Key: Orange Squares-Dwarf Starbursts, Blue Stars-Starbursts,
Red Triangles-\ulg, Green Circles-AGN \ulg
\label{S90vLFIR}}
\end{figure}

We can examine the structure of the diffuse X-ray
emission by using the images to define
its characteristic X-ray surface brightness. Specifically, we have
compared the mean surface brightness of the soft (0.3 - 1 keV) X-ray  
emission
interior to the radius enclosing
90\% of the total soft X-ray flux.
Figure \ref{S90vLFIR}
shows that this surface brightness is relatively constant over the 
almost four orders of magnitude of SFR spanned by our samples.

\begin{figure}
\includegraphics[width=6in,keepaspectratio=true,clip=true,origin=bl]{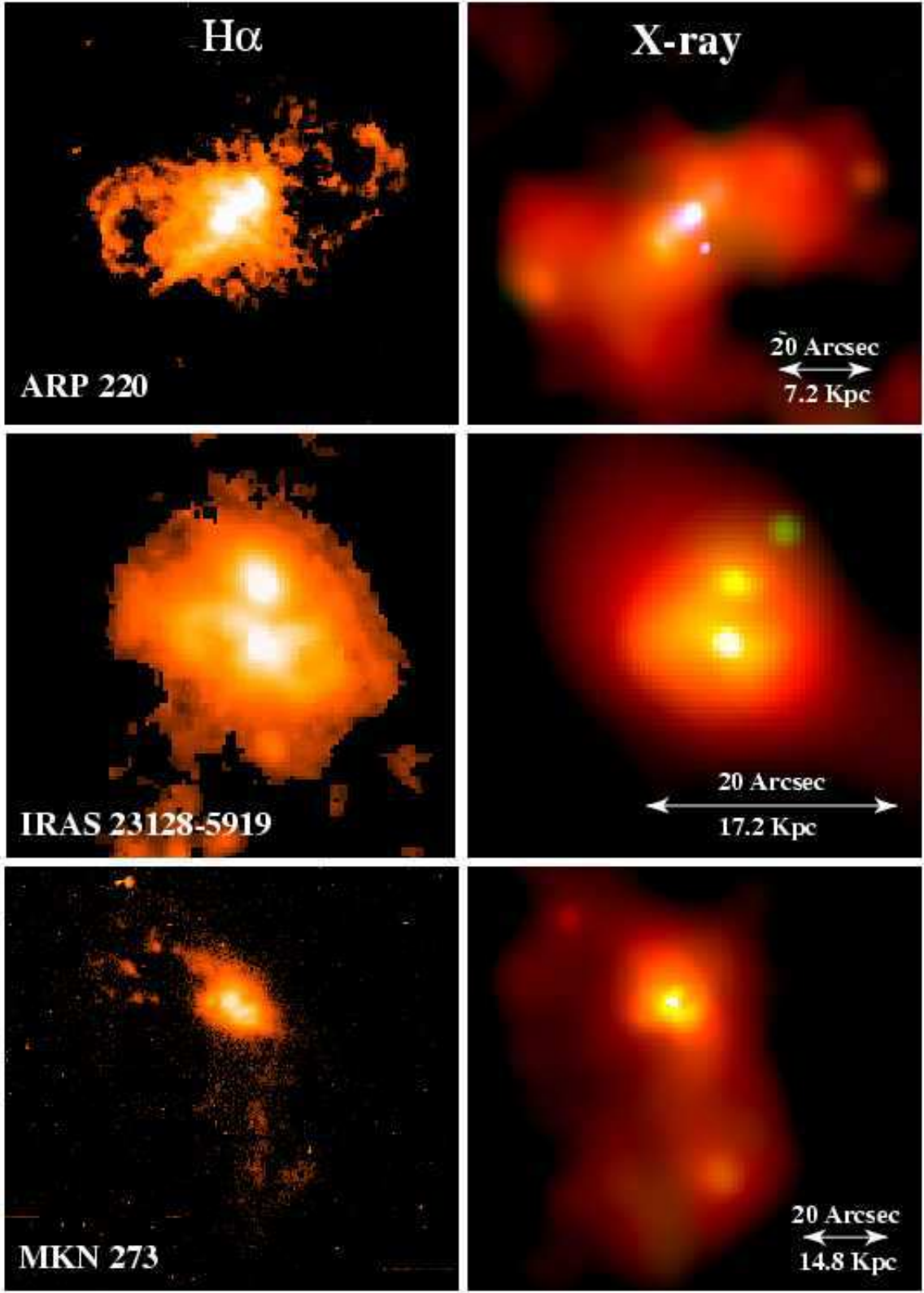}
\caption{\ha~and Adaptively Smoothed Representative Color X-ray Images 
( 0.3-1 keV Red, 1-2 keV Green, 2-8 keV Blue);
These images show a morphological relationship between the
hot gas probed by the diffuse X-ray emission and the warm gas
probed by the \ha~emission.
\label{ha}}
\end{figure}

A comparison of the diffuse X-ray and \ha~images also
shows morphological relations between these two gas phases
in all three images.  This can be seen in
in Figure \ref{ha}.  Although the X-ray images
are adaptively smoothed and have a lower
spatial resolution than the \ha~images,
there is a relationship between the
morphologies of the \ha~and diffuse X-ray emission in all three
classes of galaxies.  For every galaxy, including the ULIRGS, 
in our sample where
extended \ha~emission has been detected there is a
qualitative physical correspondence in size and shape between the
extended soft X-ray and \ha~emission. Similarly, regions lacking in
extended X-ray emission lack extended \ha~emission. 
Because of the large distances of the ULIRGs even X-ray
observations with {\it Chandra} lack the spatial resolution
to determine whether the soft X-ray and \ha~emission
are correlated (as seen in the edge-on spirals NGC 253 and NGC 3079,
\citealt{stricketal00,cecil02}) or anti-correlated (as seen in the dwarf
starburst NGC 3077, \citealt*{ott03}). Our point is merely that
the hot and warm ionized gaseous phases around these galaxies are
physically related.  In the superwind model, the
strong morphological connection between the X-ray and \ha~emission
is most likely a consequence of the hydrodynamical interaction between
the hot outflowing wind and ambient gas in the disk and halo of the  
galaxy
\citep[see][]{strick02,strick04b}.

\subsection{Scalings with the Star Formation Rate}

We now examine how the basic physical properties of the diffuse
X-ray emission (luminosity and size) scale with the fundamental
properties of the starbursts and their host galaxies (luminosity,
mass).

To do so, we have compiled information about the X-ray emission for
different spatial regions in each galaxy type.
For reference, for the \ulg~we have previously defined inner and outer  
regions.
For the edge-on starbursts
we have three regions, nucleus, disk, and halo \citep{strick04a}.
The dwarf starbursts however have only a single spatial region which
includes
the entire galaxy.  This variety of spatial regions complicates
comparison
between the galaxy types.  An additional issue is that some
regions also include powerlaw components in the X-ray spectral fits.

For simplicity, we have just added the total luminosities of
the thermal component in each region
together to get the total luminosity of the hot gas for each galaxy.
To obtain an average gas temperature we have luminosity-averaged the
temperatures of the separate spatial regions.

\begin{figure}
\centering
\leavevmode
\columnwidth=.30\columnwidth
\includegraphics[width=2in,angle=-90]{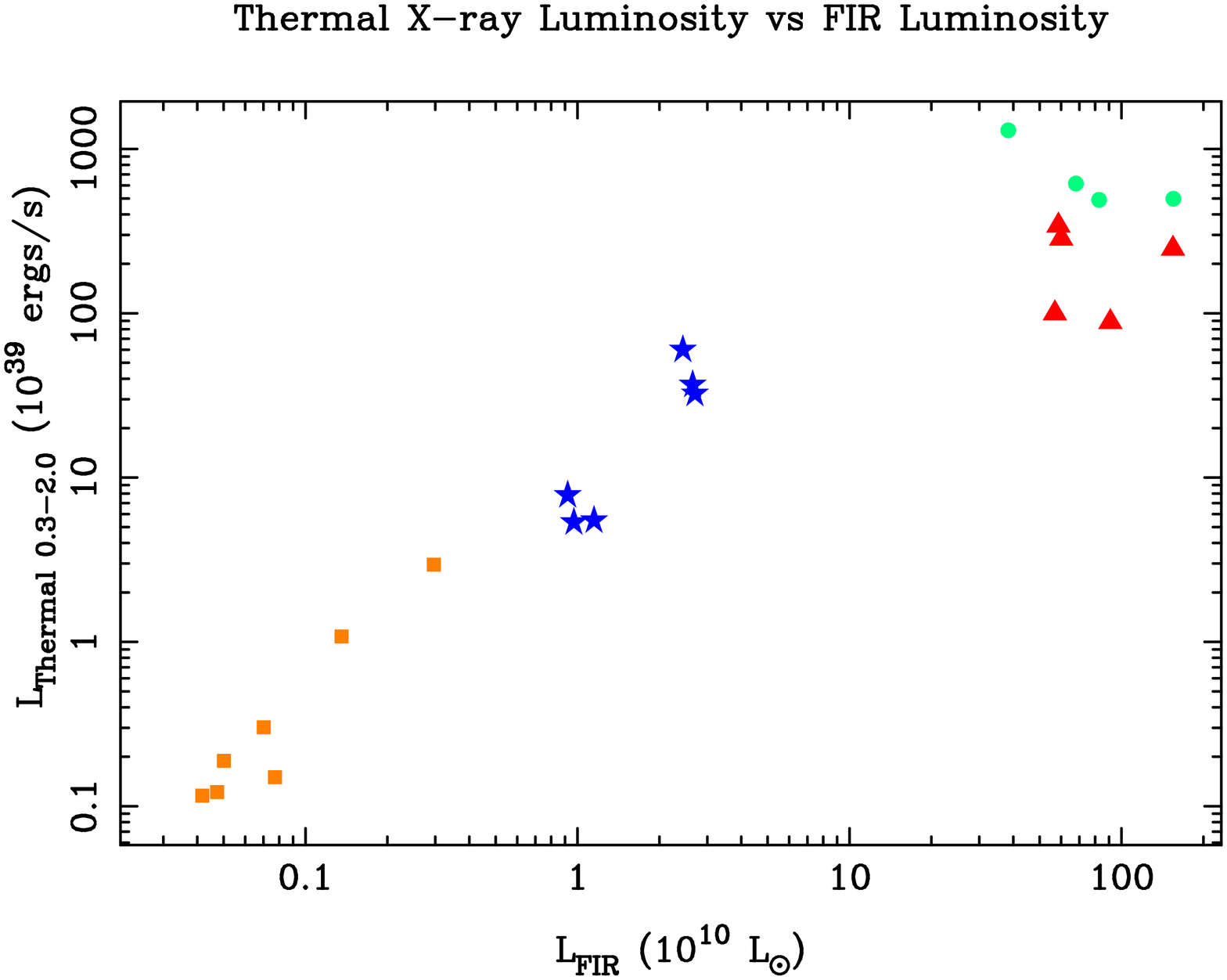}
\hfil
\includegraphics[width=2in,angle=-90]{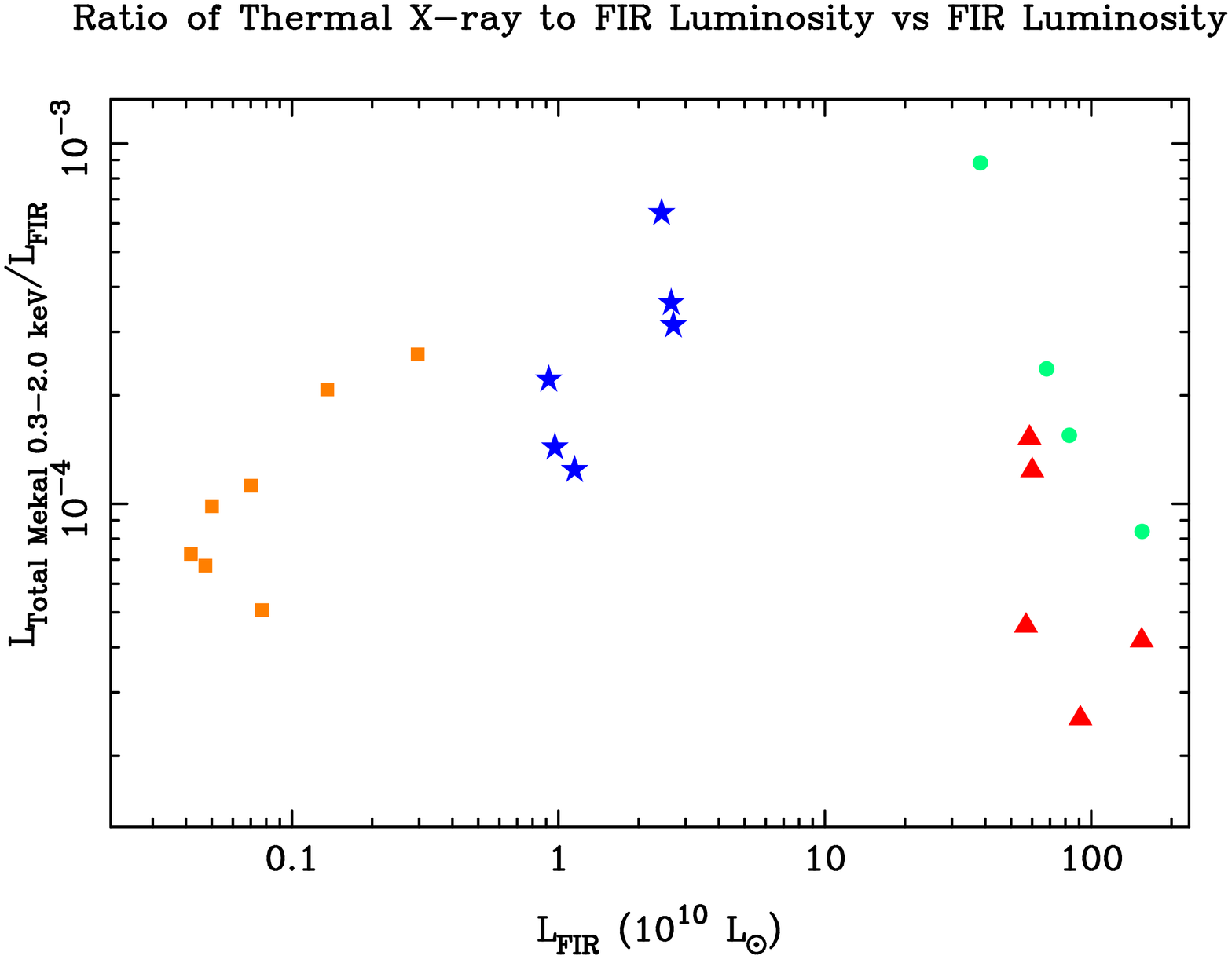}
\caption{The left shows the total thermal X-ray luminosity versus FIR  
luminosity
while the right has thermal X-ray luminosity/FIR luminosity versus FIR  
luminosity.
There is a clear correlation between
X-ray and FIR luminosity.  The plots on the right
show that the ratio of X-ray luminosity to far infrared emission is
roughly constant over almost 5 orders of magnitude.
Key: Orange Squares-Dwarf Starbursts, Blue Stars-Starbursts, Red  
Triangles-\ulg,
Green Circles-AGN ULIRGS
\label{lxrayvslfir}}
\end{figure}

In Figure \ref{lxrayvslfir} we have plotted the thermal X-ray
luminosity versus
far infrared luminosity (including the UV luminosity for the dwarfs).
As the FIR luminosity is proportional to
the star formation
rate, we see a
correlation between the thermal X-ray emission and
the star formation rate. While the good correlation between star
formation rate and X-ray luminosity has been widely noted
\citep[e.g.][]{grimm03,ran03,colb04}, our result differs from
these in that we have isolated the X-ray emission due to hot diffuse gas
(excluding the contribution to the total X-ray luminosity
from X-ray binaries and -in some cases- AGN).

On the right side of Figure \ref{lxrayvslfir} we also show
the ratio of the thermal X-ray luminosity to the far infrared
luminosity,
so that we could examine the relation more closely. Over nearly
four orders-of-magnitudes of far infrared luminosity, this ratio is
roughly constant ($\sim 10^{-4}$).  Quantitatively, we find a 
scatter of 0.38 dex for
the X-ray luminosity/FIR luminosity ratio.  The rate at which
a starburst generates mechanical energy is about 1\% of its bolometric
luminosity \citep[e.g.][]{leith95}. Figure \ref{lxrayvslfir}
thus implies that typically only about 1\% of the mechanical energy
is lost in the form of X-ray emission from hot gas. This fraction
evidently does not depend on the star formation rate itself.
Note that in Figure \ref{lxrayvslfir} we have separately color encoded
those \ulg~having AGN (based on their optical emission line properties).
See \citet{ptak03} for details.  Apart from the "nearly-a-ULIRG" NGC6240
(in which an AGN likely contributes to the soft X-ray emission),
the AGN \ulg~have ratios of thermal X-ray to FIR luminosities that
are similar to the other \ulg, edge-on starbursts, and dwarf starbursts.

\begin{figure}
\centering
\includegraphics[width=3in,angle=-90]{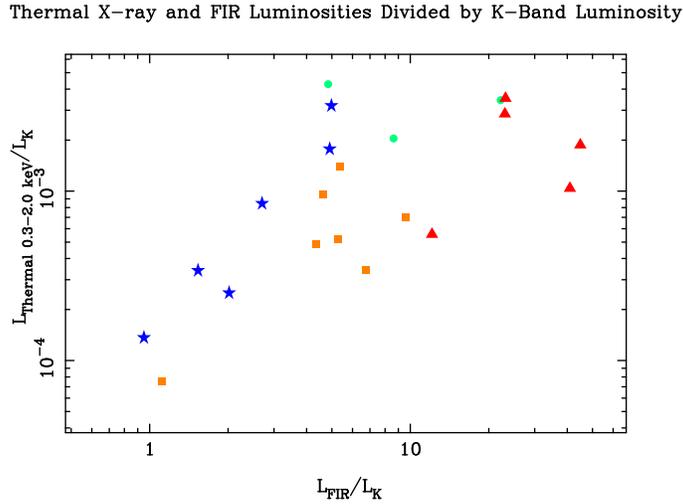}
\caption{
In this plot we have divided both the Thermal (0.3-2.0 keV)
and the FIR luminosity by the K-Band luminosity.  There is a clear
linear relation between the SFR per stellar mass and
the thermal X-ray emission per stellar mass.
Key: Orange Squares-Dwarf Starbursts, Blue Stars-Starbursts, Red  
Triangles-\ulg,
Green Circles-AGN ULIRGS
\label{div_L_K}}
\end{figure}

The correlation between FIR and thermal
X-ray luminosities is not simply an artifact of plotting luminosity
vs. luminosity (e.g. bigger galaxies have more of everything).
In Figure \ref{div_L_K} we have divided both the thermal X-ray
and FIR luminosities by the K-Band luminosity.  By renormalizing this
way, we see a clear correlation between the SFR per unit galaxy
stellar mass and the
thermal X-ray luminosity per unit mass.

\begin{figure}
\centering
\leavevmode
\columnwidth=.30\columnwidth
\includegraphics[width=2in,angle=-90]{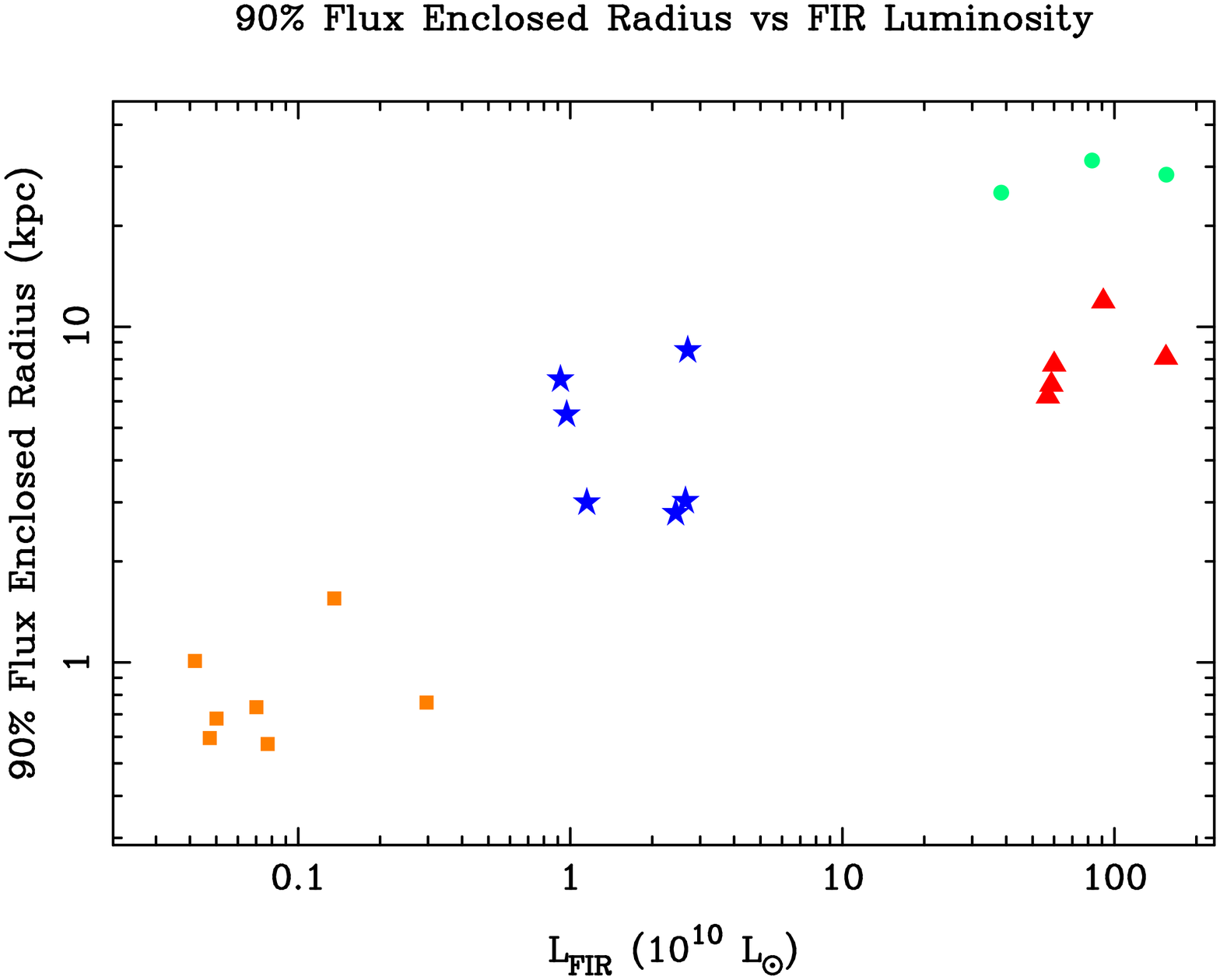}
\hfil
\includegraphics[width=2in,angle=-90]{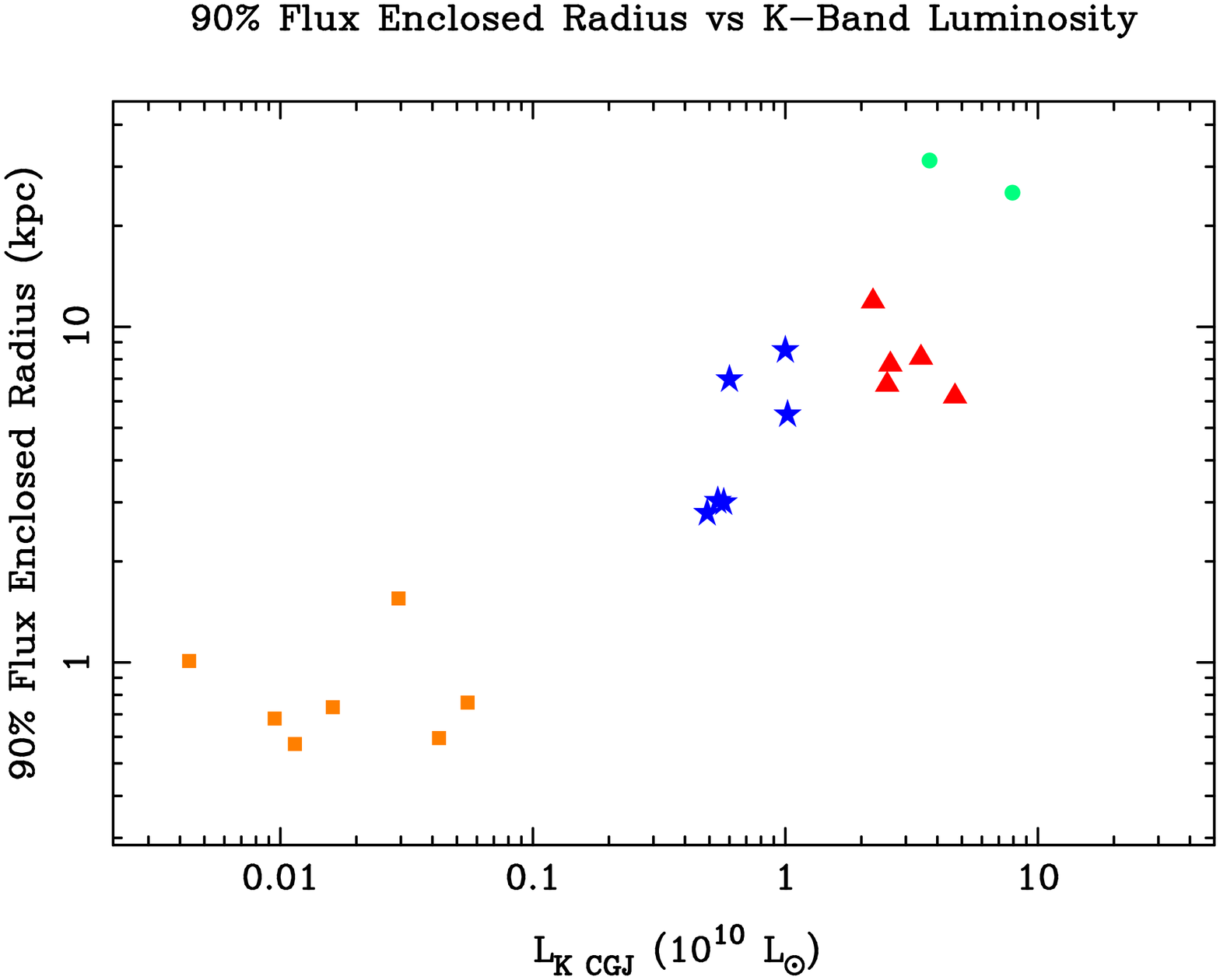}
\caption{
90\% Flux Enclosed Radii vs K-Band and FIR Luminosities.
The K-band luminosity is better correlated with the size of the
X-ray emitting region than the FIR luminosity.  This suggests that
the stellar mass determines the scale of the gas halo and not the SFR.
Key: Orange Squares-Dwarf Starbursts, Blue Stars-Starbursts,
Red Triangles-\ulg, Green Circles-AGN \ulg
\label{r90}}
\end{figure}

The size of the X-ray emitting region is also closely correlated
with both the FIR and K-band luminosities.  We have plotted
the 90\% flux enclosed radii vs luminosities in Figure \ref{r90}.
A correlation analysis using Kendall's $\tau$ rank order correlation
coefficient finds the probability of a spurious
correlation of $\rm{P}=3.4\times10^{-5}$ for the FIR to 90\% flux enclosed 
X-ray radii  and $\rm{P}=2.7\times10^{-5}$ for the K-Band.  
Therefore both the K-Band and FIR luminosity are strongly correlated to the
size of the X-ray emitting region.
The K-band relationship is marginally stronger than that in the FIR.
This relation was seen in the starburst sample by
\citet{strick04b}.  We conclude that although the host galaxy
affects the spatial extent of the X-ray emission,
the power of the X-ray emission is determined primarily by the
star formation rate.

\subsection{Chemical Composition of the Hot Gas}

The spectral degeneracy
of the $\alpha$ and Fe abundances in the vmekal models
makes it difficult to determine their absolute abundances.
However, we are able to determine \afe~ratios for the majority
of the \ulg~and dwarf starbursts (Table \ref{spectra}).
Taken as a whole we find clear evidence of enhanced \afe~ratios
relative to solar abundances.  Excluding a few measurements
with extremely large errors we find a fairly constant ratio
of about $\simeq 3$ across the sample.  We find this enrichment even
in the outer halos of the \ulg.

Similar results have been previously noticed in other starbursting
galaxies \citep[see][and references therein]{ptak97,tsuru97,mart02,fab04,strick04a}.
The enhanced \afe~ratios in the diffuse X-ray emission would be expected
if the X-ray emitting gas has been significantly enriched by the metals
returned by the starburst's supernovae. It is also possible that the
subsolar ratio is caused by significant depletion of Fe onto dust
grains in the
hot gas.

\subsection{Gas Temperature}

\begin{figure}
\centering
\leavevmode
\columnwidth=.30\columnwidth
\includegraphics[width=3in,angle=-90]{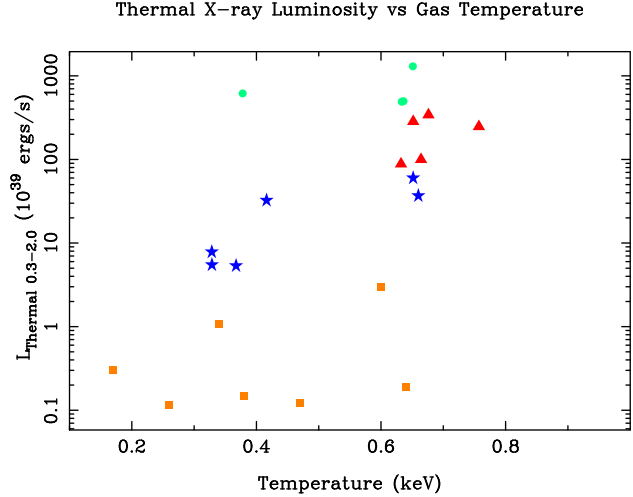}
\caption{
X-ray Temperatures vs X-ray Luminosities.  There does not appear to be  
a relation
between X-ray luminosity and temperature within our sample.
Key: Orange Squares-Dwarf Starbursts, Blue Stars-Starbursts, Red  
Triangles-\ulg,
Green Circles-AGN \ulg
\label{kT}}
\end{figure}

We have previously concluded that star formation rate
is correlated with the X-ray luminosity. This suggests that the gas mass
transported by the outflow is determined by the
mechanical energy supplied by the supernovae and stellar winds. However  
the gas
temperature might be affected by other factors.  To test this, we have  
plotted the $L_{X-ray}-T$
relation for our sample in Figure \ref{kT}. The luminosity-weighted
mean temperatures are $\sim$ 0.3 to 0.7 keV, with no strong correlation
between X-ray luminosity and gas temperature.

It does appear that the ULIRGs as a class tend to
preferentially occupy the high end of this temperature range.
There are reasons to suggest this is not caused
by contamination from the nuclear sources in the \ulg.
First, the halos of the \ulg~also have higher gas temperatures
even at large physical distances from the nuclear sources.
And second, the \ulg~with the lowest gas temperature is
\irof.  As it is the most AGN-dominated of the \ulg,
it provides a counterexample to attributing the higher gas
temperatures to AGN contamination.
Also, as the outer halos are well fit by just a thermal model
it is also probably not caused by a biased fit due to
unresolved point sources in the outer regions of the \ulg.

\begin{figure}
\centering
\includegraphics[width=2in,angle=-90]{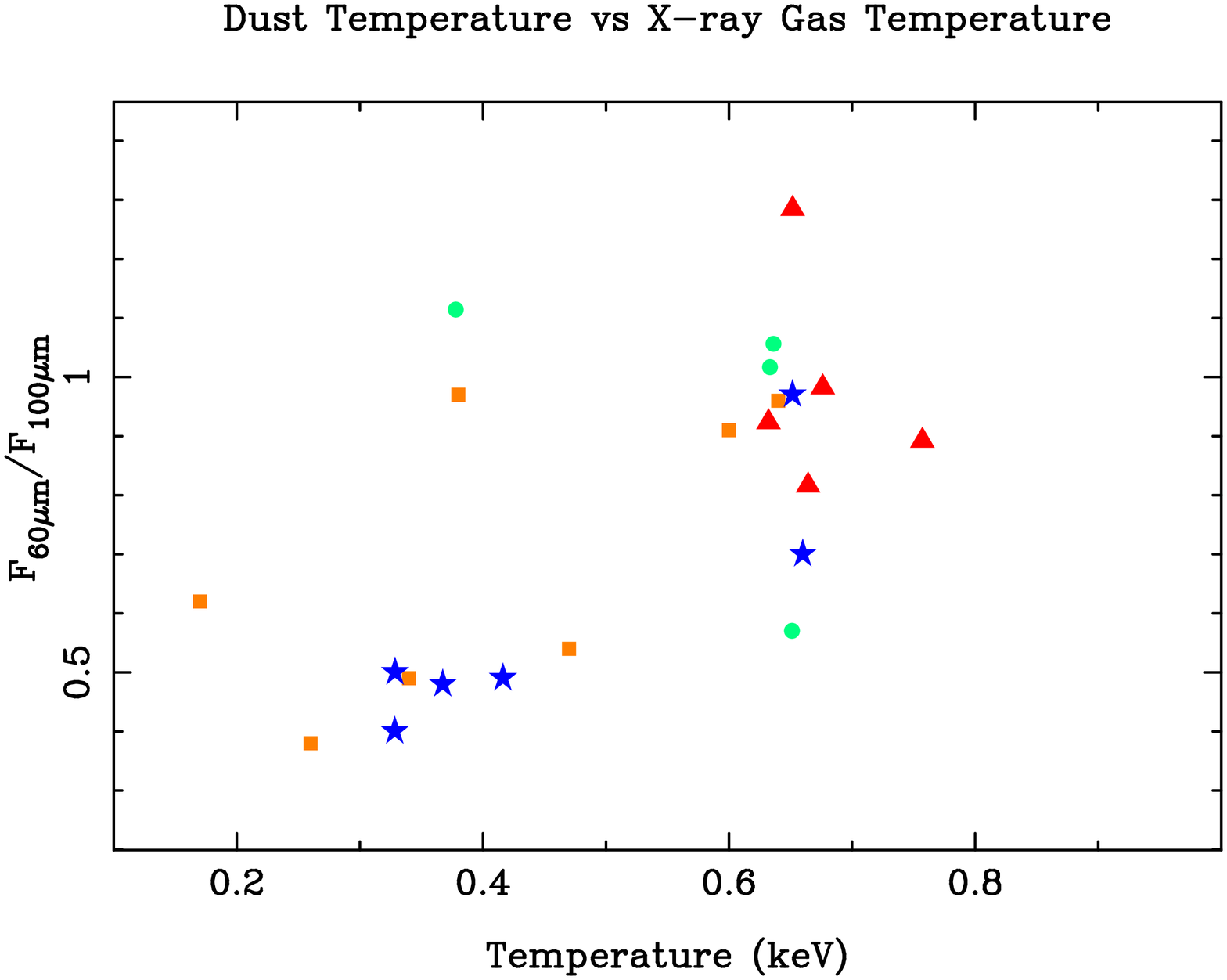}
\hfil
\includegraphics[width=2in,angle=-90]{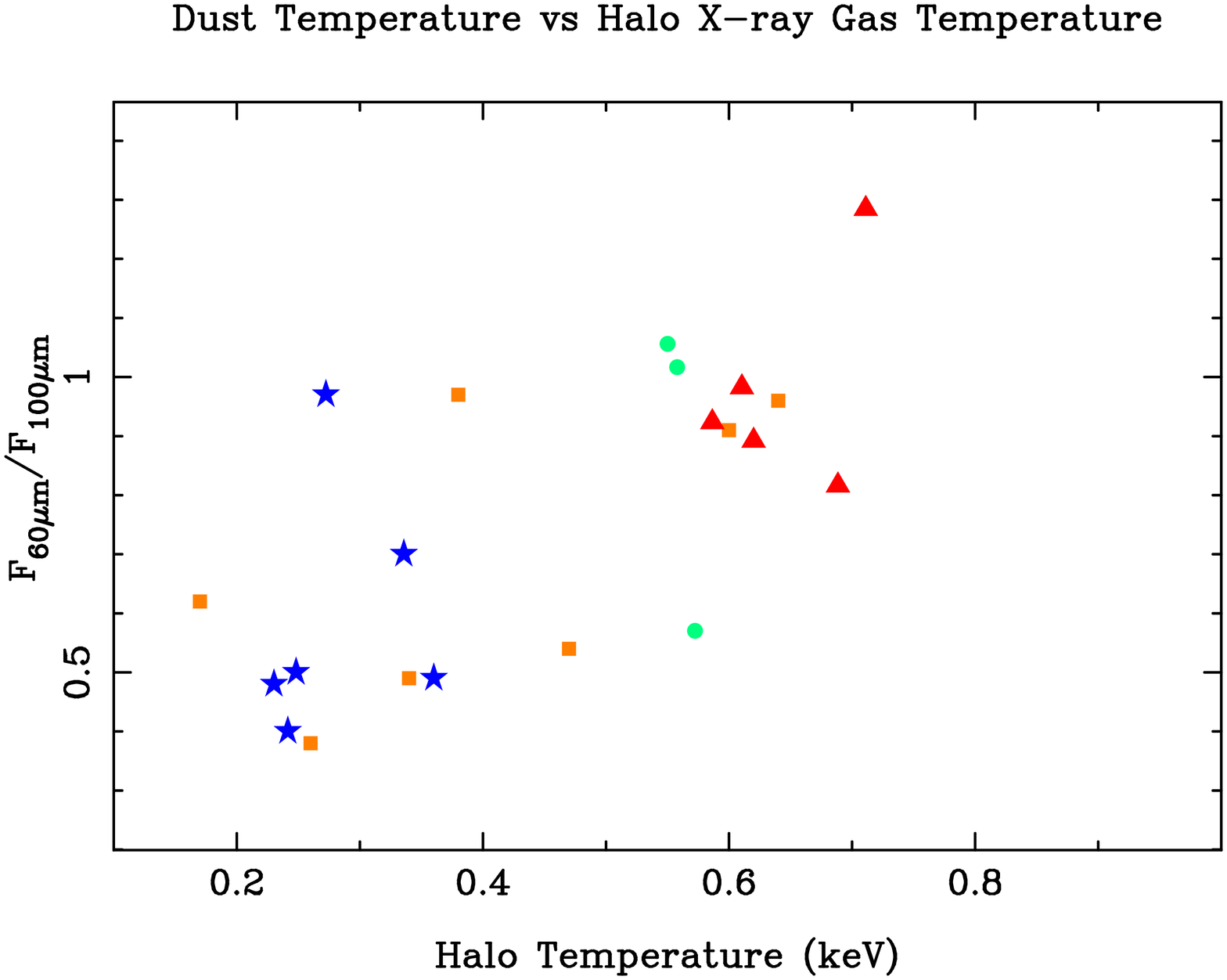}
\hfil
\caption{
$\rm{F_{60\mu m}/F_{100\mu m}}$ vs Total Mekal Temperature and Halo
Mekal Temperature.
The $\rm{F_{60\mu m}/F_{100\mu m}}$ ratio is an excellent indicator of
dust temperature. There is a slight correspondence between gas
and dust temperature.  The \ulg~however are hotter than the
dwarf starbursts and starbursts.  
Even the outer halos of the \ulg~have higher gas temperatures 
(plot on right).
Key: Orange Squares-Dwarf Starbursts, Blue Stars-Starbursts,
Red Triangles-\ulg, Green Circles-AGN \ulg
\label{F60o100vkt}}
\end{figure}

One physical difference between the ULIRGs and the other
starbursts (apart from simply their larger star formation rates),
is their larger star formation rates per unit area inside the starburst.
The star formation rate per unit area in starbursts is known to
be a key parameter in determining the observed properties of their
superwinds \citep{lehn96,strick04b}.
As shown by \citet{lehn96} the luminosity-weighted
dust temperature (as probed by the ratio of IRAS 60 and 100 $\mu$m
fluxes) correlates strongly with the star formation rate per unit area
in starbursts.  A comparison of the dust temperature and the
luminosity averaged temperatures of the X-ray emitting
gas appears in Figure \ref{F60o100vkt}.  Kendall's tau
statistic shows a rough correlation does exist
which a chance of being spurious of less than $2\%$.

\section{Conclusions and Implications}

We begin by summarizing our primary conclusions, and then briefly
discuss their implications.
Our general conclusion is that the properties of the diffuse thermal
X-ray emission in starbursts are remarkably homogeneous and follow  
simple
scaling relations over ranges of nearly four orders-of-magnitude in
X-ray luminosity and over three 
orders-of-magnitude in SFR and galaxy mass. More specifically:

\begin{itemize}

\item
The soft X-ray morphology is independent of the star formation rate,
and the same morphological relationship between the hot
X-ray emitting gas and warm optical line emission is seen
in the dwarf starbursts, the edge-on starbursts, and the ULIRGs.

\item
The X-ray luminosity is linearly proportional to the star formation rate
(estimated from the sum of the far-IR and UV luminosity). This simple
scaling holds between the star formation rate and X-ray
luminosity when both are normalized by the galaxy mass, so it is not
simply a matter of big galaxies having more of everything.

\item
The characteristic surface brightness of the diffuse thermal X-ray
emission (defined as the mean surface brightness interior to the
radius enclosing 90\% of the flux) occupies a relatively
narrow range and is independent
of the star formation rate.

\item
The characteristic size of the diffuse X-ray emission (the radius enclosing
90\% of the flux) increases systematically with increasing star
formation rate. These radii range from $\sim$0.5 to 2 kpc in the
dwarf starbursts, to 3 to 10 kpc in the edge-on starbursts, to 5 to 30
kpc in the ULIRGs.
However, the correlation of this size is slightly stronger
with K-Band luminosity (a proxy for galaxy mass) 
than with the star formation rate.

\item
The emission-weighted mean temperature of the diffuse X-ray
is $kT \sim$ 0.25 to 0.75 keV. There is a tendency for
the ULIRGs to occupy the high end of this range (above $\sim$0.6 keV).

\item
The ratio of the abundances of the $\alpha$-elements to Fe in the
diffuse gas is several times the solar value, with no dependence
on star formation rate.

\end{itemize}

These results strongly support the idea that the diffuse thermal X-ray
emission in starburst galaxies has a common physical origin. In  
particular, we
can briefly discuss the above results in the context of the model of a
galactic superwind driven by the collective energy input from supernovae
and stellar winds in the starburst.

The dynamical evolution of a starburst-driven outflow has been
extensively discussed  
\citep[e.g.][]{chev85,such94,wang95,ten98,strick00}.
Briefly, the deposition of mechanical energy by
supernovae and stellar winds results in an over-pressured cavity
of hot gas inside the starburst.
This hot gas will expand, sweep
up ambient material and thus develop a bubble-like structure.
If the ambient medium is stratified (like a disk), the superbubble will
expand most rapidly in the direction of the vertical pressure
gradient. After the superbubble size reaches several disk vertical scale
heights, the expansion will accelerate, and it is believed that
Raleigh-Taylor instabilities will then lead to the fragmentation
of the bubble's outer wall \citep[e.g.][]{mac89}.
This allows the hot gas to ``blow out''
of the disk and into the galactic halo in the form of a weakly
collimated bipolar outflow.

The strong morphological relationship between the X-ray and optical
emission in galactic winds has led to a picture in which both
trace the collision of the wind with denser ambient material in the
disk and halo of the starburst galaxy. The simplest picture is
one in which the optical line emission is produced by a shock
driven into the ambient gas by the wind, and the X-rays arise
in the ``reverse shock'' in the wind where its kinetic energy
is transformed to thermal energy and its density increases
\citep[e.g.][]{lehn99,strick02}.

This picture is consistent with the results
summarized above. The proportionality of the X-ray luminosity
and star formation rate simply means that a roughly constant fraction
($\sim$1\%) of
the kinetic energy supplied by the starburst is converted into
soft X-ray emission in the wind/cloud shocks.
\footnote{It has sometimes been suggested  - e.g. \citet{mcdo03} - that
the thermal X-ray emission in galaxy mergers like the ULIRGs
is produced by direct collisional heating of the ISM during the merger.
However, in this case, we would expect that bright thermal X-ray
emission would be a common feature of mergers of disk galaxies
(regardless of whether these mergers are undergoing a starburst).
\citet{wang97} and \citet{read98} have shown that this is
not the case. Only those
mergers with high star formation rates have correspondingly high soft  
X-ray
luminosities.}
This is consistent with the
simplest relevant theoretical model of a galactic wind, namely a  
spherically
symmetric wind-blown bubble in which the ratio of the X-ray to  
mechanical
energy input is given by
$L_X/L_{mech}\sim 0.02L^{-2/35}_{mech,42} n^{17/35}_0 t^{19/35}_7$
\citep{chu90}.  Here $L_{mech,42}$ is the rate of mechanical energy  
injection
in units of $10^{42}$ erg/s (characteristic of edge-on starbursts),
$n_0$ is the ambient particle density ($\rm{cm^{-3}}$), and $t_7$ is  
the age
of the bubble in units of $10^7$ years.

The lack of a strong
dependence of temperature on star formation rate implies that the
wind's dynamics are relatively insensitive to
the star formation rate.
Note that $k\rm{T} \approx $ 0.25 to 0.75 keV corresponds
to shock speeds of $\sim$ 450 to 800 km/s.
If these speeds are identified with that of the wind fluid, then that
implies that the wind is strongly "mass-loaded".
This is consistent with the modest
enhancements seen in the abundance ratio of the \alphe s relative to
iron, since this ratio would be boosted by the contribution from the  
core-collapse
supernovae that drive the wind \citep[e.g.][]{mart02}.

Finally, since the X-ray emission
traces the interaction between the wind and the halo/ISM, its size
should be related both the size/mass of the galaxy and to the
star formation rate (a higher star formation rate leads to a wind with
more kinetic energy which can plough its way further into the gaseous
halo of the starburst galaxy).

In summary, we believe that not only are our the results consistent  
with the
expectations for galactic superwinds, they also provide an important
benchmark against which increasingly detailed and more physically
realistic hydrodynamically simulations of galactic superwinds can be  
judged.

\newpage

\begin{deluxetable}{lccccccccc}
\tabletypesize{\footnotesize}
\tablecolumns{10}
\tablewidth{0pc}
\tablecaption{Luminosities\label{lumin}}
\tablehead{
\colhead{Galaxy} & & \multicolumn{3}{c}{$\rm{Thermal~L}_{0.3-2.0 keV}$}  
& &
\colhead{$\rm{L}_{FIR}\tablenotemark{a}$} &
\colhead{$\rm{F_{60\mu m}/F_{100\mu m}}$} & \colhead{$\rm{{L}_{K}}$} &  
\colhead{}\\
                   & & \multicolumn{3}{c}{$\rm{(ergs/s)}$}    & &
\colhead{$\rm{(L_\odot)}$} & & \colhead{$\rm{(L_\odot)}$}}
\startdata
\multicolumn{9}{c}{{\bf ULIRGs}} \\
                  & & Inner   & Outer \\
Arp 220         & & $1.6\times 10^{40}$ & $7.3\times 10^{40}$ & & &
$9.1\times 10^{11}$ & 0.92 & $2.2\times 10^{10}$ \\
IRAS 05189-2524\tablenotemark{b} & & $6.2\times 10^{41}$ & \nodata & &  
& $5.7\times
10^{11}$ & 0.82 & $4.7\times 10^{10}$ \\
IRAS 17208-0014 & & $1.8\times 10^{41}$ & $6.5\times 10^{40}$ & & &
$1.5\times 10^{12}$ & 0.89 & $3.4\times 10^{10}$ \\
IRAS 20551-4250 & & $1.7\times 10^{41}$ & $1.1\times 10^{41}$ & & &
$6.0\times 10^{11}$ & 1.28 & $2.6\times 10^{10}$ \\
IRAS 23128-5919 & & $1.7\times 10^{41}$ & $1.7\times 10^{41}$ & & &
$5.9\times 10^{11}$ & 0.98 & $2.5\times 10^{10}$ \\
MKN 231\tablenotemark{c}         & & $2.4\times 10^{41}$ & $2.6\times  
10^{41}$ & & &
$1.6\times 10^{12}$ & 1.06 & $3.3\times 10^{11}$ \\
MKN 273         & & $1.5\times 10^{41}$ & $3.4\times 10^{41}$ & & &
$8.3\times 10^{11}$ & 1.02 & $3.7\times 10^{10}$ \\
NGC 6240        & & $3.4\times 10^{41}$ & $9.5\times 10^{41}$ & & &
$3.8\times 10^{11}$ & 0.57 & $7.9\times 10^{10}$ \\
UGC05101        & & $6.8\times 10^{40}$ & $3.2\times 10^{40}$ & & &
$6.8\times 10^{11}$ & 1.11 & $7.9\times 10^{10}$
\\
\hline
\multicolumn{9}{c}{{\bf Starbursts}} \\
                  & & Nucleus & Disk    & Halo \\
NGC 253         & & $9.4\times 10^{38}$ & $3.3\times 10^{39}$ &
$1.2\times 10^{39}$ & & $1.2\times 10^{10}$ & 0.5  & $5.7\times 10^{9}$
\\
NGC 1482        & & $4.3\times 10^{39}$ & $2.8\times 10^{40}$ &
$4.3\times 10^{39}$ & & $2.7\times 10^{10}$ & 0.7  & $5.4\times 10^{9}$
\\
M 82            & & $2.8\times 10^{40}$ & $2.8\times 10^{40}$ &
$4.5\times 10^{39}$ & & $2.4\times 10^{10}$ & 0.97 & $4.9\times 10^{9}$
\\
NGC 3079        & & $1.0\times 10^{39}$ & $2.3\times 10^{40}$ &
$8.0\times 10^{39}$ & & $2.7\times 10^{10}$ & 0.49 & $1.0\times
10^{10}$ \\
NGC 3628        & & $2.4\times 10^{38}$ & $2.6\times 10^{39}$ &
$2.5\times 10^{39}$ & & $9.7\times 10^{9}$ & 0.48 & $1.0\times 10^{10}$
\\
NGC 4631        & & $6.2\times 10^{37}$ & $6.1\times 10^{39}$ &
$1.7\times 10^{39}$ & & $9.2\times 10^{9}$ & 0.4  & $6.0\times 10^{9}$
\\
\hline
\multicolumn{9}{c}{{\bf Dwarf Starbursts}} \\
                  & & All \\
He 2-10   & & $3.0\times 10^{39}$ & & & & $3.0\times 10^{9}$ & 0.91 &
$5.5\times 10^{8}$\\
NGC 1705  & & $1.2\times 10^{38}$ & & & & $4.2\times 10^{8}$ & 0.38 &
$4.3\times 10^{7}$\\
NGC 1569  & & $1.9\times 10^{38}$ & & & & $5.0\times 10^{8}$ & 0.96 &
$9.5\times 10^{7}$\\
NGC 3077  & & $1.2\times 10^{38}$ & & & & $4.7\times 10^{8}$ & 0.54 &
$4.2\times 10^{8}$\\
NGC 4214  & & $3.0\times 10^{38}$ & & & & $7.0\times 10^{8}$ & 0.62 &
$1.6\times 10^{8}$\\
NGC 4449  & & $1.1\times 10^{39}$ & & & & $1.4\times 10^{9}$ & 0.49 &
$2.9\times 10^{8}$\\
NGC 5253  & & $1.5\times 10^{38}$ & & & & $7.7\times 10^{8}$ & 0.97 &
$1.1\times 10^{8}$
\enddata
\tablenotetext{a}{Intended as an estimate of the bolometric luminosity
due to star formation. For the dwarf starburst galaxies
this value is actually the sum of the
FIR and FUV luminosities. See \S~\ref{sec:other_luminosities} for
details.}
\tablenotetext{b}{Powerlaw emission from the nuclear source dominates  
\irof 's X-ray luminosity.
This introduces large errors in calculating the thermal X-ray  
luminosity.  We detect no significant thermal emission in the outer  
region. }
\tablenotetext{c}{MKN 231's K-Band luminosity is known to be dominated  
by contamination from
the AGN.  Therefore we have not used MKN 231 in any of the plots that  
rely on its K-Band luminosity.
}
\end{deluxetable}

\begin{deluxetable}{lrllllll}
\tabletypesize{\scriptsize}
\tablecaption{Sample Properties and Observational Data
\label{obsdata}}
\tablehead{
\colhead{Galaxy} & \multicolumn{2}{c}{Chandra Pointing} &
\colhead{Distance} &
\colhead{Scale}&
\colhead{Date} & \colhead{Exposure}
& \colhead{Galactic $N_H$}\\
& \multicolumn{2}{c}{(J2000)} &  (Mpc) & \colhead{(kpc/\arcsec)} & &
\colhead{(ks)} &
\colhead{($10^{20} \rm \ cm^{-2}$)}}
\startdata
\multicolumn{8}{c}{{\bf \ulg}} \\
\\
IRAS 05189-2524 & 05 21 01.5 & -25 21 45 & 200 & 0.82 &
10/30/2001 & 5.3 & 1.93 \\
& & & & &
01/30/2002 & 13.1 \\
UGC 05101 & 09 35 51.6 & +61 21 11 & 185 &
0.76 &
05/28/2001 & 46.4 & 2.68 \\
Mkn 231 & 12 56 14.2 & +56 52 25 & 200 &
0.82 &
10/19/2000 & 25.3 & 1.25\\
Mkn 273 & 13 44 42.1 & +55 53 13 & 176 &
0.74 &
04/19/2000 & 38.8 & 1.09\\
Arp 220 & 15 34 57.1 & +23 30 11 & 84 & 0.36 &
06/24/2000 & 52.8 & 4.29\\
NGC 6240 & 16 52 58.9 & +02 24 03 & 113 & 0.49 & 07/29/2001 &
35.6 & 5.69\\
IRAS 17208-0014 & 17 23 21.9 & -00 17 00 & 200 &
0.83 &
10/24/2001 & 43.8 & 9.96\\
IRAS 20551-4250 & 20 58 26.9 & -42 39 00 & 200 &
0.83 &
10/31/2001 & 32.2 & 3.82\\
IRAS 23128-5919 & 23 15 47.0 & -59 03 17 & 210 &
0.86 &
09/30/2001 & 26.8 & 2.74\\
\hline
\multicolumn{8}{c}{{\bf Starbursts}} \\
\\
M 82      & 09 55 55.1 & +69 40 47 & 3.6  & 0.018 & 6/18/2002 & 18.0 &
4.0 \\
NGC 1482 & 03 54 39.9 & -20 30 42 & 22   & 0.107 & 2/5/2002  & 23.5 &
3.7 \\
NGC 253  & 00 47 21.3 & -25 13 00 & 2.6  & 0.013 & 12/27/1999 & 38.3 &
1.4 \\
NGC 3628 & 11 20 18.4 & +13 35 54 & 10   & 0.049 & 12/12/2000 & 54.7 &
2.2 \\
NGC 3079 & 10 01 53.5 & +55 40 42 & 17   & 0.083 & 3/7/2001 & 26.5 &
0.8 \\
NGC 4631 & 12 41 56.9 & +32 37 34 & 7.5  & 0.036 & 4/16/2000 & 55.7 &
1.3 \\
\hline
\multicolumn{8}{c}{{\bf Dwarf Starbursts}} \\
\\
He 2-10 &  08 36 15.2 & -26 24 34 & 9   & 0.044 & 3/23/2001 & 17.6 &
9.7  \\
NGC 1569 & 04 30 49.0 & +64 50 54 & 2.2 & 0.010 & 4/11/2000 & 78.7 &
22.6   \\
NGC 1705 & 04 54 13.7 & -53 21 41 & 5.1 & 0.025 & 9/12/2003 & 32.1 &
3.85  \\
NGC 3077 & 10 03 21.0 & +68 44 02 & 3.6 & 0.017 & 3/08/2001 & 45.5 &
3.9    \\
NGC 4214 & 12 15 38.7 & +36 19 42 & 2.9 & 0.014 & 10/16/2001 & 26.4 &
1.5 \\
NGC 4449 & 12 28 12.0 & +44 05 41 & 2.9 & 0.014 & 2/05/2001 & 21.2 &
1.5    \\
NGC 5253 & 13 39 56.0 & -31 38 24 & 3.2 & 0.015 & 1/14/2001 &  39.9 &
3.9
\enddata
\tablecomments{Positions and redshifts
were obtained from NED.  Galactic $N_H$ values are determined from
   using the COLDEN tool}
\end{deluxetable}

\begin{deluxetable}{lrrccc}
\tabletypesize{\footnotesize}
\tablecolumns{6}
\tablewidth{0pc}
\tablecaption{Inner Region Definitions for \ulg \label{innerregion}}
\tablehead{
\colhead{Galaxy} & \colhead{RA} & \colhead{DEC} & \colhead{Major Axis}
          & \colhead{Minor Axis} & \colhead{Rotation} \\
                  & \colhead{(J2000)} & \colhead{(J2000)}
                  & \colhead{(Arcsec)} & \colhead{(Arcsec)} &  
\colhead{90-PA}}
\startdata
ARP~220 & +15:34:57.254 & +23:30:11.56 & 7.72 & 5.23 & 9.95 \\
IRAS~05189-2524 & +05:21:01.393 & -25:21:45.37 & 3.52 & 3.52 & 0 \\
IRAS~17208-0014 & +17:23:21.984 & -00:17:00.38 & 8.46 & 4.55 & 118.28 \\
IRAS~20551-4250 & +20:58:26.778 & -42:39:00.19 & 5.16 & 5.16 & 0 \\
IRAS~23128-5919 & +23:15:46.725 & -59:03:15.50 & 3.42 & 3.42 & 0 \\
MKN~231                 & +12:56:14.206 & +56:52:25.33 & 3.34 & 3.34 &  0 \\
MKN~273                 & +13:44:42.054 & +55:53:12.73 & 4.75 & 3.74 & 319.58 \\
NGC~6240              & +16:52:58.897 & +02:24:03.62 & 10.8 & 7.67 &  103.14 \\
UGC~05101            & +09:35:51.602 & +61:21:11.98 & 4.28 & 4.28 & 0
\enddata
\end{deluxetable}

\begin{deluxetable}{lccccccccccc}
\rotate
\tabletypesize{\footnotesize}
\tablecolumns{12}
\tablewidth{0pc}
\tablecaption{Enclosed Light and Isophotal Radii
\label{obsradii}
}
\tablehead{
\colhead{Galaxy} &  \colhead{Scale} &
\colhead{$r_{50\%}$\tablenotemark{a}}  &  
\colhead{$r_{75\%}$\tablenotemark{a}}  &  
\colhead{$r_{90\%}$\tablenotemark{a}}  &
\colhead{$r_{95\%}$\tablenotemark{a}}
& \colhead{$\Sigma_{50\%}$\tablenotemark{b}}  &  
\colhead{$\Sigma_{75\%}$\tablenotemark{b}}  &
\colhead{$\Sigma_{90\%}$\tablenotemark{b}}  &  
\colhead{$\Sigma_{95\%}$\tablenotemark{b}}
& \colhead{$r_{1e-09}$\tablenotemark{c}} &  
\colhead{$\frac{L_{\rm{Thermal~X-ray}}}{\pi r_{90\%}^2}$}\\
   & \colhead{$\rm{kpc}/\arcsec$} & \multicolumn{4}{c}{kpc}
& \multicolumn{4}{c}{$\rm{photons}~\rm{s}^{-1}~\rm{cm}^{  
-2}~\rm{arcsec}^{-2}$}
& \colhead{kpc}  & \colhead{$10^{38} \rm{ergs/s/kpc^2}$}}
\startdata
ARP 220 & 0.39 &   3.90 & 8.32 & 11.9 & 13.6 &   2.56e-08 &
1.12e-08 & 7.17e-09 & 5.88e-09 & $ 15.7$ & 1.99\\

IRAS 05189-2524\tablenotemark{d} & 0.89 &    0.17 & 1.13 & 1.71 & 3.48 &
8.07e-07 & 8.07e-07 & 8.07e-07 & 5.19e-07 & $ 10.7$ & 672\\

IRAS 17208-0014 & 0.89 &    2.84 & 5.40 & 8.09 & 9.10 &
4.08e-08 & 2.39e-08 & 1.40e-08 & 1.17e-08 & $ 11.8$ & 12.0\\

IRAS 20551-4250 & 0.89 &    1.32 & 4.35 & 7.72 & 10.8 &
1.38e-07 & 9.64e-08 & 4.52e-08 & 2.59e-08 & $ 15.3$ & 15.2\\

IRAS 23128-5919 & 0.93 &    2.11 & 3.72 & 6.72 & 8.26 &
1.49e-07 & 1.03e-07 & 5.12e-08 & 2.72e-08 & $ 13.5$ & 24.1\\

MKN 231 & 0.89 &    3.63 & 16.0 & 28.4 & 35.9 &   8.97e-08 &
2.07e-08 & 8.87e-09 & 6.11e-09 & $ 33.9$ & 1.97\\

MKN 273 & 0.79 &    3.12 & 19.8 & 31.3 & 39.9 &   1.31e-07 &
1.79e-08 & 9.44e-09 & 6.36e-09 & $ 38.7$ & 1.59\\

NGC 6240 & 0.52 &    4.71 & 14.3 & 25.1 & 32.2 &   2.72e-07 &
6.07e-08 & 2.60e-08 & 1.71e-08 & $ 41.5$ & 6.56\\

UGC05101 & 0.83 &    0.93 & 3.24 & 6.20 & 8.75 &   5.95e-08 &
4.81e-08 & 1.97e-08 & 1.34e-08 & $ 11.8$ & 8.28\\
\hline
\multicolumn{12}{c}{Starbursts} \\
M 82 ACIS-S & .0175 &    0.989 & 1.78 & 2.78 & 3.55 &
9.82e-08 & 6.00e-08 & 3.53e-08 & 2.53e-08 & $ 5.49$ & 24.6\\
NGC 1482 & .1071 &    0.927 & 1.74 & 3.02 & 3.69 &   1.45e-07
& 7.96e-08 & 3.56e-08 & 2.60e-08 & $ 4.84$ & 12.8\\
NGC 253 ACIS-S & .0126 &    0.822 & 1.88 & 2.99 & 3.70 &
3.53e-08 & 1.50e-08 & 8.82e-09 & 7.13e-09 & $ 4.57$ & 1.94\\
NGC 3079 & .0829 &    2.71 & 6.05 & 8.50 & 10.0 &   2.20e-08 &
8.66e-09 & 5.68e-09 & 4.50e-09 & $ 10.1$ & 1.42\\
NGC 3628 & .0485 &    3.38 & 4.60 & 5.47 & 5.90 &   2.81e-09 &
2.39e-09 & 2.10e-09 & 1.92e-09 & $ 5.73$ & 0.57\\
NGC 4631 & .0364 &    3.55 & 5.64 & 6.96 & 7.56 &   4.40e-09 &
2.92e-09 & 2.39e-09 & 2.16e-09 & $ 7.35$ & 0.51\\
\hline
\multicolumn{12}{c}{Dwarf Starbursts} \\

He 2-10 & 0.0435 &    0.182 & 0.490 & 0.759 & 0.906 &
4.02e-07 & 1.13e-07 & 6.00e-08 & 4.52e-08 & $ 1.29$ & 16.3\\
NGC 1569 & 0.0107 &    0.297 & 0.509 & 0.680 & 0.739 &
1.08e-08 & 5.84e-09 & 4.03e-09 & 3.62e-09 & $ 0.783$ & 1.30\\
NGC 1705 & 0.0247 &    0.538 & 0.870 & 1.01 & 1.06 &
1.16e-09 & 7.47e-10 & 6.90e-10 & 6.71e-10 & $ 0.482$ & 0.36\\
NGC 3077 & 0.0174 &    0.231 & 0.384 & 0.595 & 0.701 &
1.37e-08 & 8.26e-09 & 4.42e-09 & 3.43e-09 & $ 0.553$ & 1.10\\
NGC 4214 & 0.0142 &    0.271 & 0.549 & 0.735 & 0.841 &
1.43e-08 & 6.57e-09 & 4.74e-09 & 4.08e-09 & $ 0.945$ & 1.78\\
NGC 4449 & 0.0142 &    0.711 & 1.10 & 1.55 & 1.77 &   1.42e-08
& 9.50e-09 & 5.88e-09 & 4.80e-09 & $ 1.82$ & 1.43\\
NGC 5253 & 0.0152 &    0.151 & 0.356 & 0.571 & 0.676 &
4.14e-08 & 1.30e-08 & 6.41e-09 & 4.94e-09 & $ 0.650$ & 1.46\\
\enddata
\tablenotetext{a}{Radius enclosing 50\%, 75\%, 90\%, or 95\% of the
counts in the 0.3-1.0 keV X-ray bandpass.}
\tablenotetext{b}{Average surface brightness inside the radius
enclosing 50\%, 75\%, 90\%, or 95\% of the counts in the 
0.3-1.0 keV X-ray bandpass.}
\tablenotetext{c}{Radius at the $1\times10^{-9}~\rm{counts/s/cm^2}$  
surface brightness.}
\tablenotetext{d}{The essentially pointlike nature of the X-ray
emission of \irof~does not allow us to reliably
determine any spatial information.  Therefore we have omitted
\irof~from our radii and surface brightness plots.}
\end{deluxetable}

\begin{deluxetable}{lcccrrrrr}
\rotate
\tabletypesize{\footnotesize}
\tablecolumns{9}
\tablewidth{0pc}
\tablecaption{Spectral Fits \label{spectra}}
\tablehead{
\colhead{Galaxy} & \colhead{Region}   & \colhead{kT}    & \colhead{$K$}
&
\colhead{$\alpha/\rm{Fe}$} & \colhead{$N_{H}$} & \colhead{PL
Normalization}   & \colhead{$\Gamma$}
& \colhead{$\chi^2/\rm{DOF}$}\\
\colhead{} & \colhead{}   & \colhead{(keV)}    & \colhead{} &
\colhead{} &
\colhead{($10^{22}~\rm{cm^{-2}}$)}   & \colhead{}    & \colhead{} &
\colhead{}}
\startdata
ARP~220 & Inner & $0.85^{+0.05}_{-0.05}$ & $3.1^{+7.5}_{-1.9}\times
10^{-6}$ &
$7.6^{+7.8}_{-3.0}$
& $0.5^{+0.3}_{-0.2}$ & $1.4^{+0.7}_{-0.4}\times 10^{-5}$ &
$1.3^{+0.2}_{-0.3}$ & 18.4/23\\
   & Outer &  $0.59^{+0.03}_{-0.04}$ & $1.4^{+0.3}_{-0.3}\times10^{-4}$ &
$2.6^{+0.7}_{-1.0}$ &
& & & 39.3/37\\
IRAS~05189-2524\tablenotemark{a,b} & Inner & $0.38^{+0.04}_{-0.03}$ &
$6.0^{+1.2}_{-1.2}\times 10^{-4}$ &
\nodata{} & $3.8^{+0.7}_{-0.1}$ & $3.0^{+1.6}_{-0.2}\times 10^{-4}$ &
$0.9^{+0.3}_{-0.1}$ & 128/100\\
   & Outer & \nodata{} & \nodata{} & \nodata{} \\
IRAS~17208-0014\tablenotemark{b} & Inner & $0.81^{+0.22}_{-0.19}$ &
$6.3^{+31.4}_{-2.2}\times 10^{-5}$ &
\nodata{} & $3.1^{+4.8}_{-1.9}$ & $2.2^{+9.7}_{-2.2}\times 10^{-5}$ &
$1.8^{+2.3}_{-0.6}$
& 5.6/9\\
   & Outer &  $0.62$ & $6.6\times10^{-5}$ & \nodata{} &
& & & 1.1/0\\
IRAS~20551-4250 & Inner & $0.61^{+0.04}_{-0.04}$ &
$7.4^{+46}_{-5.1}\times 10^{-6}$ &
$2.8^{+1.5}_{-1.9}$
& $0.1^{+0.6}_{-0.1}$ & $4.6^{+4.1}_{-2.2}\times 10^{-6}$ &
$1.0^{+0.3}_{-0.4}$ & 17.1/12\\
   & Outer &  $0.71^{+0.06}_{-0.06}$ & $2.8\times10^{-6}$ &
$5.4^{+2.9}_{-2.3}$ &
& & & 6.0/4\\
IRAS~23128-5919 & Inner & $0.74^{+0.10}_{-0.09}$ &
$2.0^{+1.8}_{-1.9}\times 10^{-5}$ &
$10.0^{+6.7}_{-4.8}$
& $7.8^{+1.3}_{-2.9}$ & $3.2^{+15}_{-2.4}\times 10^{-4}$ &
$2.8^{+0.5}_{-0.4}$ & 12.5/11\\
   & Outer &  $0.61^{+0.07}_{-0.05}$ & $1.6\times10^{-5}$ &
$6.5^{+4.6}_{-2.8}$ &
& & & 3.3/4\\
MKN~231 & Inner & $0.73^{+0.13}_{-0.09}$ & $6.6^{+2.4}_{-0.7}\times
10^{-5}$ &
$5.3^{+4.6}_{-1.1}$
& $0.7^{+0.3}_{-0.4}$ & $2.4^{+0.3}_{-0.6}\times 10^{-5}$ &
$0.4^{+0.2}_{-0.1}$ & 30.1/40\\
   & Outer &  $0.55^{+0.04}_{-0.04}$ & $1.4^{+0.3}_{-0.4}\times10^{-4}$ &
$3.6^{+1.9}_{-1.1}$ &
& & & 16.4/18\\
MKN~273 & Inner & $0.80^{+0.06}_{-0.04}$ & $2.8^{+4.5}_{-1.6}\times
10^{-6}$ &
$3.9^{+2.1}_{-1.8}$
& $39.2^{+3.3}_{-4.6}$ & $4.8^{+1.5}_{-1.5}\times 10^{-4}$ &
$1.5^{+0.2}_{-0.3}$ & 61.3/53\\
   & Outer &  $0.56^{+0.06}_{-0.06}$ & $1.3^{+0.5}_{-0.5}\times10^{-4}$ &
$3.1^{+1.0}_{-0.7}$ &
& & & 39.5/36\\
NGC~6240 & Inner & $0.87^{+0.05}_{-0.04}$ & $9.6^{+43.8}_{-2.3}\times
10^{-6}$ &
$3.6^{+0.7}_{-0.5}$
& $0.4^{+0.1}_{-0.1}$ & $3.3^{+0.4}_{-0.4}\times 10^{-4}$ &
$1.8^{+0.1}_{-0.1}$ & 272/177\\
   & Outer &  $0.57^{+0.02}_{-0.02}$ & $9.2^{+1.4}_{-2.0}\times10^{-4}$ &
$2.8^{+0.8}_{-0.5}$ &
& & & 99.4/84\\
UGC~05101\tablenotemark{b} & Inner & $0.65^{+0.08}_{-0.10}$ &  
$2.1^{+11}_{-1.3}\times
10^{-7}$ &
$10.2^{+24.5}_{-7.4}$
& $0.9^{+1.2}_{-0.7}$ & $1.1^{+1.7}_{-0.5}\times 10^{-5}$ &
$1.4^{+0.7}_{-0.5}$ & 10.2/9\\
   & Outer &  $0.69^{+0.14}_{-0.11}$ & $1.1\times10^{-5}$ & \nodata{} &
& & & 0.3/0\\
\hline
\multicolumn{9}{c}{{\bf Dwarf Starbursts}} \\
\\
He 2-10\tablenotemark{c} & All &  $0.60^{+0.03}_{-0.03}$ &  
$3.1^{+1.0}_{-0.8}\times
10^{-4}$ &
                   $2.6^{+0.7}_{-0.6}$ & $4.3^{+2.3}_{-2.0}$ &
                   $8.4^{+43}_{-5.7}\times10^{-5}$ & $1.4^{+0.8}_{-0.3}$
& 40/41  \\
NGC 1569\tablenotemark{d} & All &
                   $0.64^{+0.02}_{-0.02}$ & $6.2^{+1.0}_{-0.7}\times
10^{-4}$ &
                   $2.4^{+0.4}_{-0.4}$  & \nodata &
                   \nodata & \nodata & 111/83  \\

NGC 1705\tablenotemark{b,d} & All &
                   $0.26^{+0.03}_{-0.04}$ & $8.4\times 10^{-5}$ &
                   \nodata  & \nodata &
                  \nodata & \nodata & 9.2/5  \\

NGC 3077\tablenotemark{d} & All & $0.47^{+0.05}_{-0.04}$ & $1.5\times  
10^{-4}$ &
                   $1.7^{+0.9}_{-0.8}$ &  \nodata &
                   \nodata & \nodata & 31/20\\

NGC 4214\tablenotemark{e} & All & $0.17$ & \nodata & \nodata & \nodata
& \nodata
                   & \nodata & \nodata\\

NGC 4449\tablenotemark{c} & All &
                   $0.33^{+0.03}_{-0.02}$ & $1.9^{+0.4}_{-1.4}\times
10^{-3}$ &
                   $3.0^{+1.3}_{-0.8}$ & $0.3^{+0.3}_{-0.3}$ &
                   $4.1^{+1.7}_{-0.8}\times10^{-4}$ & $2.2^{+0.2}_{-0.2}$
& 182/133 \\

NGC 5253\tablenotemark{d} & All &
                   $0.39^{+0.02}_{-0.03}$ & $9.4\times 10^{-5}$ &
                   $2.5^{+0.6}_{-0.5}$ &  \nodata &
                   \nodata & \nodata & 56/39

\enddata
\tablenotetext{a}{We were unable to detect any significant X-ray
emission in the outer region of \irof~as the source is point like in  
the X-ray.}
\tablenotetext{b}{There were not enough counts in the spectra to
determine the \afe~ratio for these observations.}
\tablenotetext{c}{These dwarf starburst galaxies, like the inner
regions of the \ulg, required a powerlaw to obtain and acceptable fit.
The redshifted $N_H$ column absorbs only the powerlaw and is in  
addition to
the galactic $N_H$ in Table \ref{obsdata}.}
\tablenotetext{d}{These starbursts did not require a powerlaw component  
or an additional $N_H$ column to obtain an acceptable fit.}
\tablenotetext{e}{\citet{hart04}}
\end{deluxetable}


\clearpage

\acknowledgments

We are grateful to Chris Mihos for sending us the \ha~images
of \irtt~and to J.
Hartwell for providing the X-ray data for NGC 4214.

\end{document}